%% file: main.tex
\documentclass[twocolumn]{aastex631}

\newcommand{\jybm}{Jy~beam$^{-1}$}

\begin{document}

\title{The JCMT BISTRO Survey:
The Magnetic Fields of the IC 348 Star-forming Region}

\input{authorlist}

%\firstauthor{Youngwoo Choi}
%\email{cyw3614@snu.ac.kr}
\correspondingauthor{Woojin Kwon}
\email{wkwon@snu.ac.kr}

\begin{abstract}

We present 850 $\mu$m polarization observations of the IC 348 star-forming region in the Perseus molecular cloud as part of the B-fields In STar-forming Region Observation (BISTRO) survey. We study the magnetic properties of two cores (HH 211 MMS and IC 348 MMS) and a filamentary structure of IC 348. We find that the overall field tends to be more perpendicular than parallel to the filamentary structure of the region. The polarization fraction decreases with intensity, and we estimate the trend by power-law and the mean of the Rice distribution fittings. The power indices for the cores are much smaller than 1, indicative of possible grain growth to micron size in the cores. We also measure the magnetic field strengths of the two cores and the filamentary area separately by applying the Davis-Chandrasekhar-Fermi method and its alternative version for compressed medium. The estimated mass-to-flux ratios are 0.45-2.20 and 0.63-2.76 for HH 211 MMS and IC 348 MMS, respectively, while the ratios for the filament is 0.33-1.50. This result may suggest that the transition from subcritical to supercritical conditions occurs at the core scale ($\sim$ 0.05 pc) in the region. In addition, we study the energy balance of the cores and find that the relative strength of turbulence to the magnetic field tends to be stronger for IC 348 MMS than HH 211 MMS. The result could potentially explain the different configurations inside the two cores: a single protostellar system in HH 211 MMS and multiple protostars in IC 348 MMS.

\end{abstract}


\section{Introduction} \label{sec:intro}

Molecular clouds are the densest and coldest regions of the interstellar medium where new stars are born. Recent observations toward nearby molecular clouds made by $Herschel$ revealed that molecular clouds have highly filamentary structures \citep[e.g.,][]{Andr__2010,Arzoumanian_2011,Palmeirim_2013}. Molecular cores lie along filaments, forming new stars by gravitational collapse \citep[e.g.,][]{Hacar_2013,Fernandez_2014,Pattle_2023}.

One of the crucial recent findings is that these filamentary structures are strongly affected by local magnetic fields \citep[e.g.,][]{Palmeirim_2013,Cox_2016,Planck_2016,Ward_Thompson_2017,Pattle_2023}. The magnetic fields are perpendicular to high column density filaments, while they are parallel to lower dense sub-filaments or striations \citep[e.g.,][]{Cox_2016,Liu_2018,Alina_2019,Fissel_2019,Soam_2019,Doi_2020,Pillai_2020,Arzoumanian_2021,Kwon_2022,Ching_2022,Ward-Thompson_2023}. The observed field morphology can be explained by the magnetic funneling scenario, in which diffuse matter moves through sub-filaments along the field lines, feeding materials to the main filaments \citep{Soler_2013,Andr__2014,Pattle_2023}.

Other prominent features of molecular clouds are that they are significantly hierarchical \citep{Pokhrel_2018} and have a lifetime longer than the free-fall timescale. Large clouds fragment into small clumps or filamentary systems and form smaller cores inside a filament. This whole process is known to be far slower than the free-fall collapse of clouds, given that star formation efficiency in nearby molecular clouds is estimated to be less than 5$\%$ \citep[e.g.,][]{Carpenter_2000,Evans_2009,Barnes_2017}. This indicates that the star formation process is not only regulated by gravity and thermal pressure but also affected by other processes like magnetic fields and turbulence \citep[e.g.,][]{Padoan_2014}. The relative importance of these four properties determines star formation efficiency and how molecular clouds fragment into smaller structures \citep{Tang_2019,Chung_2023}. Therefore, estimating the relative strengths between the energy densities of gravity, thermal pressure, turbulence, and magnetic fields on different scales in molecular clouds is vital to understanding the star formation process \citep[e.g.,][]{Pattle_2017,Liu_2019,Coude_2019,Wang_2019,Lyo_2021,Hwang_2021,Hwang_2022,Chung_2023,Tahani_2023}.

Magnetic properties of molecular clouds can be studied through dust polarization observations. Non-spherical grains can be spun-up to suprathermal by radiative torques and become aligned with their shortest axis parallel to the ambient magnetic fields \citep{Draine_1997,Lazarian_2007}.  This grain alignment process includes the alignment of the grain shortest axis with its angular momentum (internal alignment) caused by internal relaxation processes \citep{Purcell_1979}, and the alignment of the grain angular momentum with the magnetic field (external alignment) caused by radiative torques and magnetic relaxation  \citep[e.g.,][]{Hoang_2016}. As a result, continuum emission from dust grains is linearly polarized along the direction of the grain longer axis, which is perpendicular to the magnetic field. This allows us to infer the magnetic field orientations projected on the plane of the sky by rotating 90$^{\circ}$ the linear polarization angles observed at far-IR to (sub-)millimeter wavelengths \citep[e.g.,][]{Matthews_2009}. Note that, in contrast, the background radiation at optical and near-infrared wavelengths is extincted by aligned dust grains, resulting in polarization parallel to the magnetic field directions \citep{Davis_1951}.

The Perseus giant molecular cloud complex is one of the largest and nearest sites in which numerous young stars are born. It contains various star-forming molecular clouds including IC 348, NGC 1333, B1, L 1448, and L 1455 \citep{Kirk_2007}. Hundreds of young stars are born in the Perseus giant molecular cloud \citep{Bally_2008}, and most belong to two clusters, IC 348 and NGC 1333 \citep{Herbst_2008, Walawender_2008,Luhman_2016}. IC 348 is located in the eastern part of the Perseus molecular cloud complex. The distance of IC 348 is estimated to be around 321 pc from trigonometric parallax observations \citep{Ortiz_Le_n_2018} and 295 pc from a combination of stellar photometry, parallax measurement, and CO observations \citep{Zucker_2018}. We assumed a distance of 300 pc, with an uncertainty of 10\%. IC 348 contains a large number of pre-main sequence stars and protostars \citep{Herbst_2008}, and the age of IC 348 is from 2 to 6 Myr based on evolutionary models \citep{Luhman_2003, Bell_2013}. Most pre-main sequence stars are concentrated within the central region, whereas protostars lie in the southwestern ridge part \citep{Herbst_2008}. In this paper, we focus on the southwestern part of IC 348, also known as IC 348-SW. Two cores (HH 211 MMS and IC 348 MMS) host Class 0 protostars in IC 348-SW, and we study the magnetic properties of the filamentary structure and the two cores.

This study was conducted as part of the BISTRO (B-Fields in Star-forming Region Observations) survey \citep{Ward_Thompson_2017}, one of the James Clerk Maxwell Telescope (JCMT) large programs. The BISTRO survey observes molecular clouds of the Gould Belt \citep{Herschel_1847, Gould_1879} to investigate the role of magnetic fields in star formation \citep{Ward_Thompson_2017} and has been extended to BISTRO-2 and 3 for further and/or massive star-forming regions. IC 348 is a target of BISTRO-2.

In this paper, we present 850 $\mu$m polarization observations toward the IC 348 star-forming region. Observations and data reduction are described in Section \ref{sec:observations}. In Section \ref{sec:magnetic field morphology}, we show the magnetic field morphology of IC 348. We also discuss the dependence of the polarization fraction on total intensity by power-law and the mean of the Rice distribution fittings in Section \ref{sec:intensity and polarization fraction}. In Section \ref{sec:magnetic field strength}, we estimate the magnetic field strengths of the two cores and the filament. In Section \ref{sec:energy calculations}, we study the energy balances of the two core regions. We discuss the aligned grain size, magnetic criticality, and stability of the target region in Section \ref{sec:discussion}. Finally, we summarize our conclusions in Section \ref{sec:conclusion}.

\section{Observations} \label{sec:observations}

JCMT is a 15 m single-dish radio telescope located at an altitude of 4092 m in Mauna Kea, Hawaii. The Submillimeter Common User Bolometer Array-2 (SCUBA-2) mounted on JCMT has a 10,000-pixel submillimeter continuum and simultaneously observes in the 450 $\mu$m and the 850 $\mu$m atmospheric windows \citep{Holland_2013}. Furthermore, to obtain polarization data, the POL-2 polarimeter is installed in the pathway of the SCUBA-2 camera and observes the sky in a daisy-like pattern at a speed of 8$''$ $\mathrm{s}^{-1}$ \citep{Friberg_2016}.

We observed the IC 348 star-forming region with POL-2 between October 2019 and February 2020 as part of the BISTRO survey (Project ID: M17BL011). Observations of 20 repetitions were made for 40 minutes each, to reach a total of about 14 hours under Band 1 weather conditions ($ \tau_{225\mathrm{GHz}} < 0.05$), using the POL-2 DAISY scan mode \citep{Friberg_2016}. In addition to BISTRO observations, 16 sets of polarization data (30 minutes each) were taken between July (Project ID: M17AP073) and September (Project ID: M17BP058) 2017 under Band 2 weather conditions ($0.05 < \tau_{225\mathrm{GHz}} < 0.08$) (PI: Woojin Kwon). This additional data were combined with BISTRO observations. The rms noises of the final Stokes $I$ and Stokes $Q$ and $U$ maps at 850 $\mu$m are 3.3 mJy beam$^{-1}$ and 2.5 mJy beam$^{-1}$, respectively.

The observational data were reduced using the $pol2map$ command of the SMURF package of STARLINK \citep{Currie_2014}. This reduction process comprises three steps. First, the $calcqu$ command creates Stokes $I$, $Q$, and $U$ timestreams from the raw bolometer timestreams for each observation. Then, the initial Stokes $I$ map is created by co-adding $I$ timestreams using the iterative map-making routine $makemap$ \citep{Chapin_2013}. In the next stage, the pointing corrections and SNR-based fixed mask determined from the initial $I$ map are applied to re-produce a Stokes $I$ map. In this step, instead of the $makemap$ routine, the $skyloop$ command is used to generate a combined map from all the observations. In the last step, the Stokes $Q$ and $U$ maps are produced using the output masks created in the second stage. The final Stokes $I$ map is used for instrumental polarization ($IP$) correction of the final $Q$ and $U$ maps.  We use the `August 2019' $IP$ model\footnote{\url{https://www.eaobservatory.org/jcmt/2019/08/new-ip-models-for-pol2-data/}} for $IP$ correction. The pixel size of the final $I$, $Q$, and $U$ maps is set to 4$''$. Lastly, the final half-vector catalog of polarizations is made from the final $I$, $Q$, and $U$ maps on a bin size of 12$''$, similar to the beam size of SCUBA-2 at 850 $\mu$m, 14.6$''$.

In this study, we use the 850 $\mu$m $I$, $Q$, and $U$ maps to infer the magnetic properties of the IC 348 star-forming region. The $I$ map at 450 $\mu$m is also used to calculate the dust temperature and column density distributions of the target with the 850 $\mu$m $I$ map. The effective full-widths at half-maximum (FWHM) of the SCUBA-2 beams are 14.6$''$ and 9.8$''$ at 850 $\mu$m and 450 $\mu$m, respectively \citep{Dempsey_2013}. The Flux Conversion Factors (FCF) of SCUBA-2 at 850 $\mu$m are 495 Jy pW$^{-1}$ beam$^{-1}$ post 2018 June and 516 Jy pW$^{-1}$ beam$^{-1}$ from 2016 November to 2018 June. At 450 $\mu$m, the values are 472 Jy pW$^{-1}$ beam$^{-1}$ post 2018 June and 531 Jy pW$^{-1}$ beam$^{-1}$ pre 2018 June \citep{Mairs_2021}. Weighted by the observational time, we adopt 503 Jy pW$^{-1}$ beam$^{-1}$ at 850 $\mu$m, 495 Jy pW$^{-1}$ beam$^{-1}$ at 450 $\mu$m for FCF values of SCUBA-2. These values were multiplied by transmission corrector factors for POL-2, which are 1.35 and 1.96 at 850 $\mu$m and 450 $\mu$m, respectively \citep{Friberg_2016}. Therefore, in this paper, we apply the FCF values of 679 Jy pW$^{-1}$ beam$^{-1}$ at 850 $\mu$m and 970 Jy pW$^{-1}$ beam$^{-1}$ at 450 $\mu$m.

We also use the C$^{18}$O $J = 3 \rightarrow 2$ data at the rest frequency of 329.278 GHz to infer turbulent motions of the target region, which were taken under the project code M06BGT02 and first published by \cite{Curtis_2010} using the reduction performed by the JCMT Gould Belt Survey \citep{Ward_Thompson_2007}. The survey observed nearby low-mass and intermediate-mass star-forming regions including IC 348 with SCUBA-2 and the Heterodyne Array Receiver Program (HARP). HARP offers large-scale velocity maps of a high spectral resolution with 16 heterodyne pixels \citep{Buckle_2009}. The pixel size and spectral resolution of the map are 7$''$ and 0.1 km s$^{-1}$, respectively. The rms noise of the spectra is 0.22 K.

\section{Magnetic Field Morphology} \label{sec:magnetic field morphology}

\begin{figure*}[ht!]
\epsscale{0.8}
\plotone{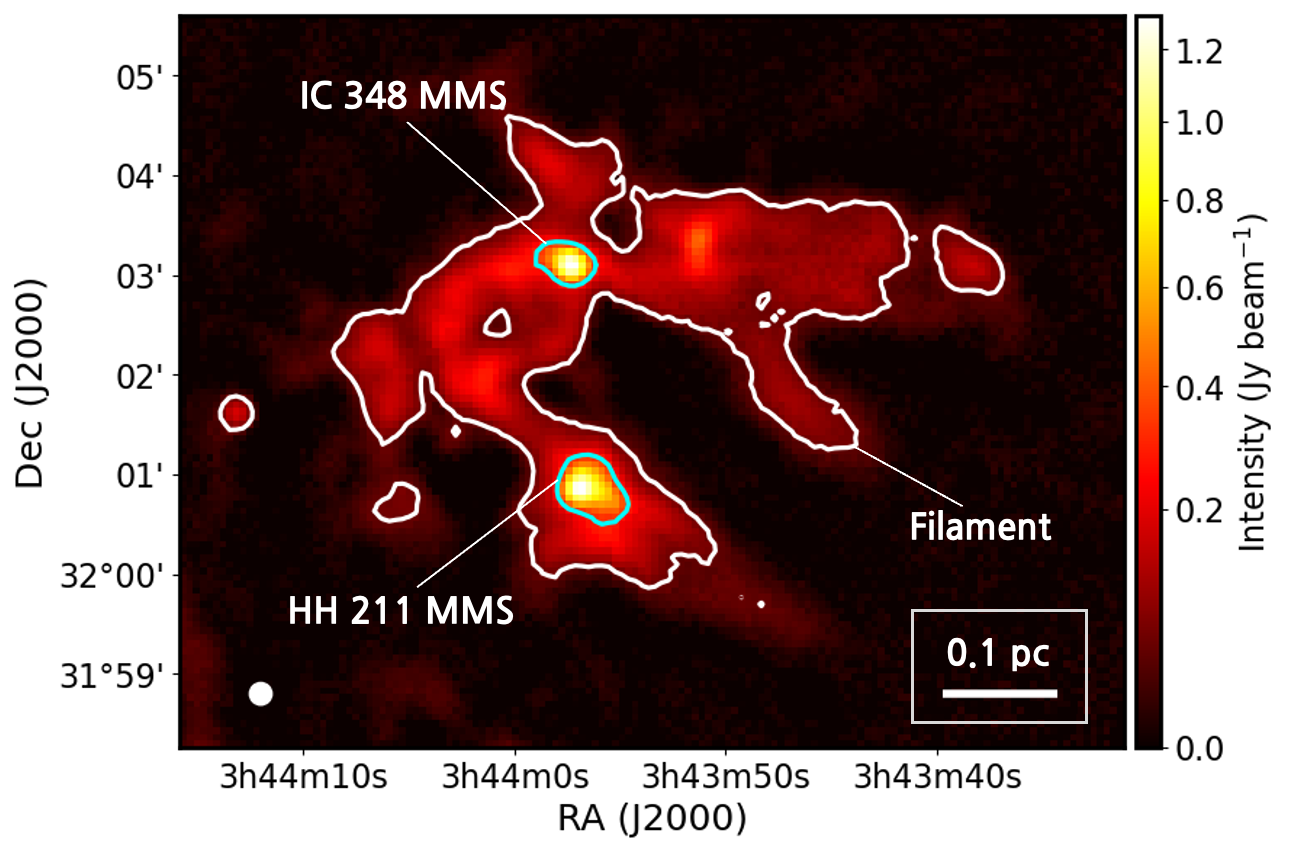}
\caption{Positions of HH 211 MMS and IC 348 MMS (or IC 348 SMM2) in the IC 348 star-forming region at the total intensity map of 850 $\mu$m. Cyan and white contours are at total intensities of 0.3 \jybm and 0.05 \jybm, respectively, representing the core and filament regions. The beam and the spatial scale bar are shown in the lower left and right corners, respectively.
\label{fig:positions}}
\end{figure*}

Figure \ref{fig:positions} shows two core regions and a filamentary structure of the IC 348 star-forming region. We define the area where 850 $\mu$m intensities are greater than 0.3 \jybm (90$\sigma$) as the core region and 0.05 \jybm (15$\sigma$) as the filament region.

\begin{figure*}[ht!]
\epsscale{1.15}
\plottwo{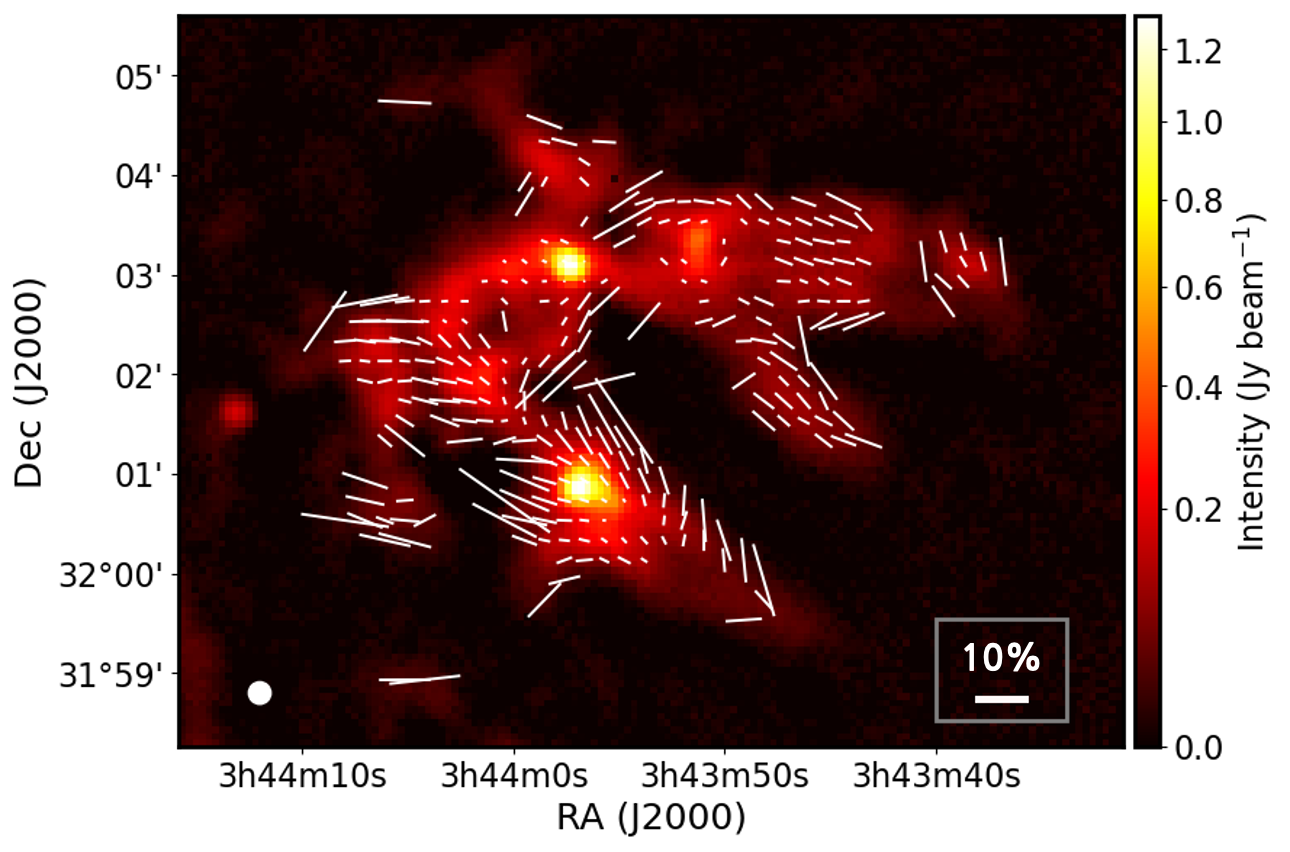}{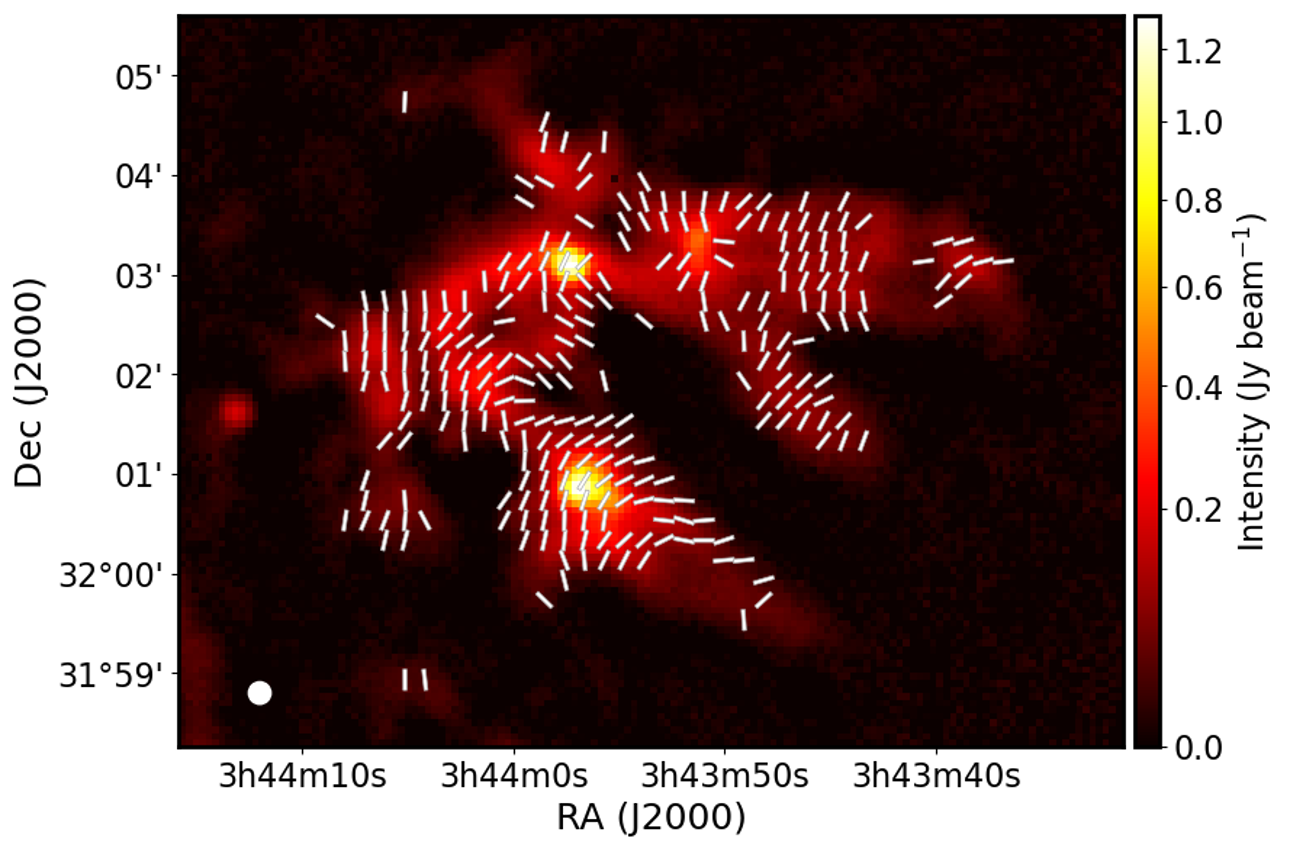}
\caption{Left: Linear polarization angles and fractions in the IC 348 star-forming region detected in 850 $\mu$m dust continuum. The length of each segment is proportional to the polarization fraction, and the reference length is shown at the right bottom. Right: Magnetic field orientations in the IC 348 star-forming region obtained by 90° rotations of the linear polarization angles. Segment lengths are uniform to have the field directions better presented. The background image is the 850 $\mu$m total intensity, and the beam size is shown at the bottom left of both panels.
\label{fig:vector}}
\end{figure*}

\begin{figure*}[ht!]
\epsscale{1.15}
\plotone{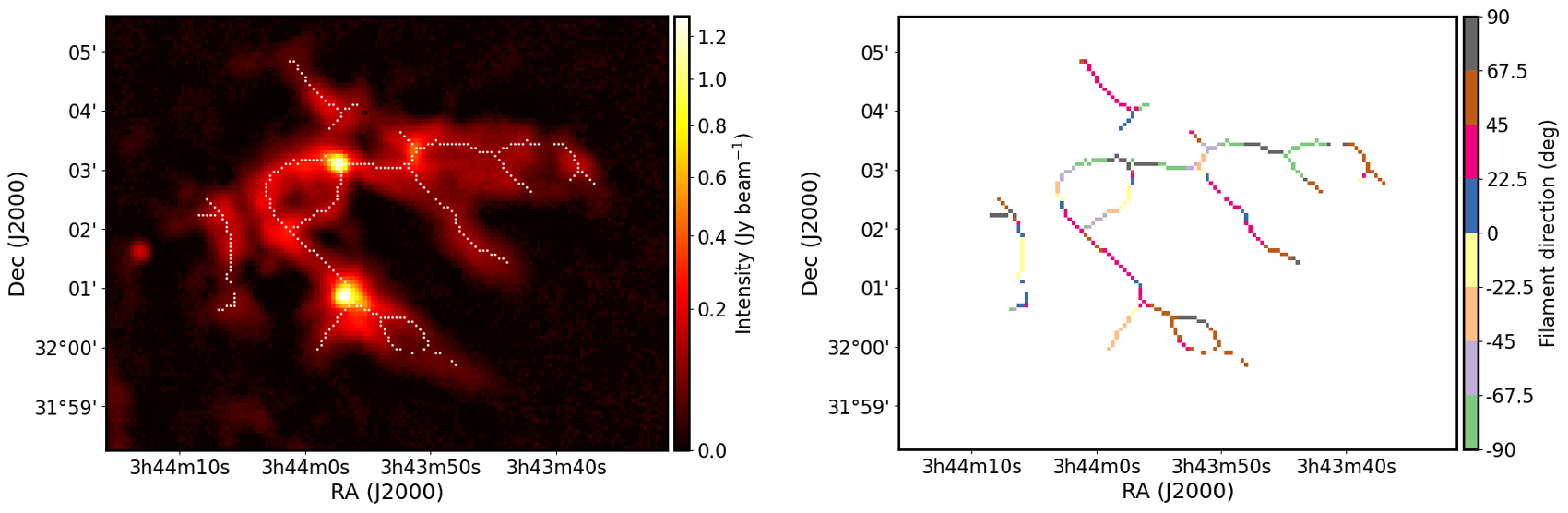}
\caption{Left: Filament skeleton of the IC 348 star-forming region. Right: Position angles of the filament at individual positions of the filament skeleton.
\label{fig:fil_direc}}
\end{figure*}

\begin{figure*}[ht!]
\epsscale{1.15}
\plotone{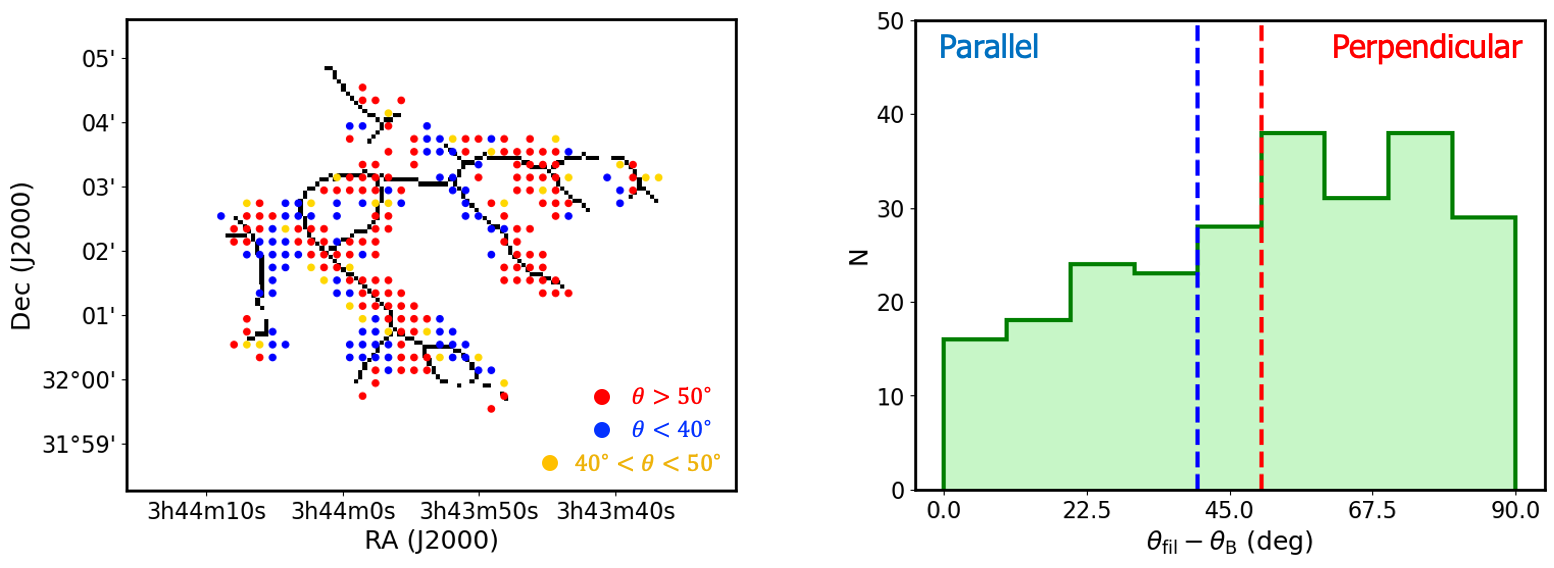}
\caption{Left: The relation between the filament and magnetic fields at each location of the half-vectors. Red dots represent perpendicular-like points ($\theta > 50$\textdegree), while blue dots represent parallel-like points ($\theta < 40$\textdegree). Yellow dots represent neither perpendicular-like nor parallel-like (40\textdegree $< \theta <$ 50\textdegree). Right: Histogram of the measured angles between the filament and magnetic fields of the IC 348 star-forming region. Blue and red dashed lines represent 40 and 50 degrees, respectively.
\label{fig:fil_ang}}
\end{figure*}

Polarization properties are inferred from the Stokes $I$, $Q$, and $U$ maps obtained from the data reduction process described in the previous section. In principle, from the $Q$ and $U$ map values, polarized intensity ($PI$) is calculated as
\begin{equation}
PI_{\mathrm{non-debiased}} = \sqrt{Q^2 + U^2},
\end{equation}
where $Q$ and $U$ represent Stokes $Q$ and $U$ values, respectively.
However, since the polarized intensity is defined as a sum of squared Stokes $Q$ and $U$ values, errors of the Stokes $Q$ and $U$ lead to an overestimation in the measured polarized intensity. This positive bias is usually corrected using the uncertainty in the $Q$ and $U$ values:
\begin{equation}
PI_{\mathrm{debiased}} = \sqrt{Q^2 + U^2 - 0.5(\delta Q^2 + \delta U^2)},
\end{equation}
where $\delta Q$ and $\delta U$ represent uncertainties in the Stokes $Q$ and $U$, respectively \citep{Wardle_1974}. The uncertainty of polarized intensity is given by
\begin{equation}
\delta PI = \sqrt{\frac{Q^2 \delta Q^2 + U^2 \delta U^2}{Q^2 + U^2}}.
\end{equation}
The polarization fraction and its uncertainty are
\begin{equation}
P = \frac{PI}{I}
\end{equation}
and
\begin{equation}
\delta P = \sqrt{\frac{\delta PI^2}{I^2} + \frac{\delta I^2(Q^2+U^2)}{I^4}},
\end{equation}
where $I$ and $\delta I$ represent a Stokes $I$ value and its uncertainty, respectively.

Lastly, the polarization angle and its uncertainty are calculated as follows:
\begin{equation}
\theta_{\mathrm{p}} = \frac{1}{2}\mathrm{arctan}\left(\frac{U}{Q}\right)
\label{eq:theta}
\end{equation}
and
\begin{equation}
\delta\theta_{\mathrm{p}} = \frac{1}{2} \frac{\sqrt{Q^2\delta U^2 + U^2 \delta Q^2}}{Q^2+U^2}.
\end{equation}

The left panel of Figure \ref{fig:vector} shows 850 $\mu$m polarization angles within the IC 348 star-forming region. Polarization angles are obtained from the Stokes $Q$ and $U$ values using Equation \ref{eq:theta} and sampled on a 12$''$ grid, similar to the beam size of the observations. Half vectors are plotted where (1) $I/\delta I>10$, (2) $PI/\delta PI>3$ and (3) $\delta P<5\%$. The length of each vector indicates its polarization fraction. The right panel of Figure \ref{fig:vector} shows the magnetic field morphology of the IC 348 star-forming region. Magnetic field orientations are inferred by rotating the observed polarization angles by 90°, assuming that the major axis of non-spherical dust grains is perpendicular to local magnetic fields \citep[e.g.,][]{Andersson_2015}.

To study the relationship between magnetic fields and the filamentary structure of the region, we measure angles between magnetic field orientations and filament directions. We first found a filament skeleton using the $filfinder$ package \citep{Koch_2015} with a size threshold of 200 square pixels. The extracted filament skeleton is shown in the left panel of Figure \ref{fig:fil_direc}. Then we derived the Hessian matrix of the 850 $\mu$m intensity map to determine the filament directions in each skeleton pixel. Since the Hessian matrix is a second-order partial derivative of a scalar function, it provides information about the curvature of the scalar function near any given point. Thus, by deriving the Hessian matrix of the intensity field, the two eigenvectors at any given position are directly related to the intensity contrast at that point. Before deriving the Hessian matrix, we convolved the intensity map with a Gaussian 2D kernel with a standard deviation of 16$''$ to make a smoothed map, since smoothing through a Gaussian function of an order of the instrumental beam effectively suppresses pixel-to-pixel noise levels while minimizing blurring of the filamentary structures \citep{Schisano_2014}. A pair of eigenvectors of the Hessian matrix at skeleton pixels are derived, and the filament direction is accepted to be the eigenvector direction with the smaller eigenvalue. For detailed descriptions of a filament direction identification, refer to \cite{Schisano_2014}. The right panel of Figure \ref{fig:fil_direc} shows the filament directions at each skeleton pixel. The displayed angles are measured positively from north to east in the equatorial frame, ranging from -90 to 90 degrees. Then, we matched observed magnetic field segments with the nearest filament skeleton pixels and measured the angle between the magnetic field orientation of the segment and the filament direction of the nearest pixel. In this process, the angles were accepted when the distance between a magnetic field segment and the nearest skeleton pixel is smaller than 30$''$, which corresponds to about 0.043 pc.

The relative orientations between the filament and magnetic fields at the locations of observed magnetic field segments can be seen in the left panel of Figure \ref{fig:fil_ang}. Red dots represent the points where the measured angle is larger than 50° (perpendicular-like). In contrast, blue dots represent where the measured angle is smaller than 40° (parallel-like). Yellow dots are the points between 40° and 50°. We applied rough criteria for identifying perpendicular-like and parallel-like data points, given that the measured angles are based on plane-of-sky projections. For example, even if the magnetic field and a filament are orthogonal in three-dimensional space, the projected angles between them could be significantly less than 90°, depending on the inclination : see Section 5.2 of \cite{Doi_2020}. The right panel of Figure \ref{fig:fil_ang} shows the distribution of the measured angles between the filament and magnetic field of the region. There are more cases close to the perpendicular orientation than those to the parallel one.

\begin{figure}[ht!]
\epsscale{1.15}
\plotone{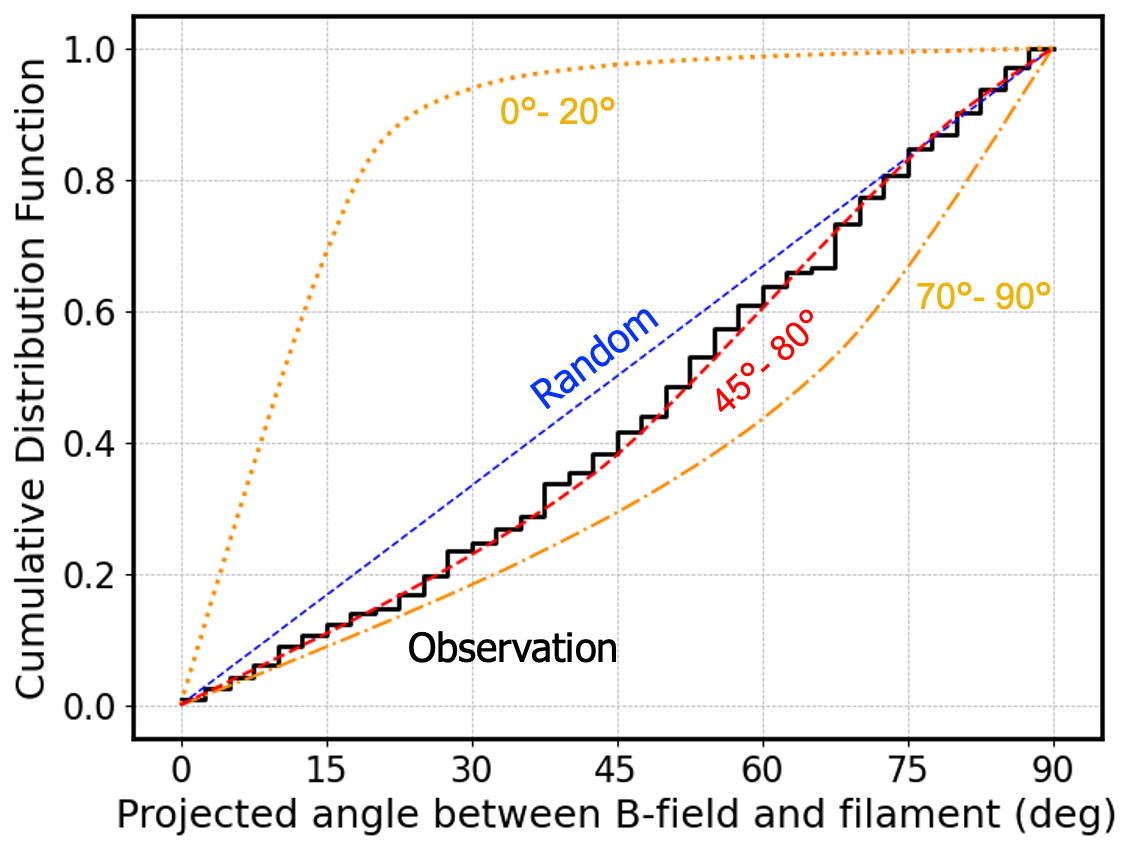}
\caption{Cumulative distribution function of projected angles between magnetic fields and filaments. The black line shows the CDF of observed projected angles between magnetic fields and the filament. Orange lines indicate expected CDFs when three-dimensional angles are from 0° to 20° and from 70° to 90°, respectively. The blue line represents when magnetic fields and filaments are randomly oriented. The red line shows the qualitatively best-fit case (45°-80°).
\label{fig:cdf}}
\end{figure}

\begin{figure*}[ht!]
\epsscale{1.15}
\plottwo{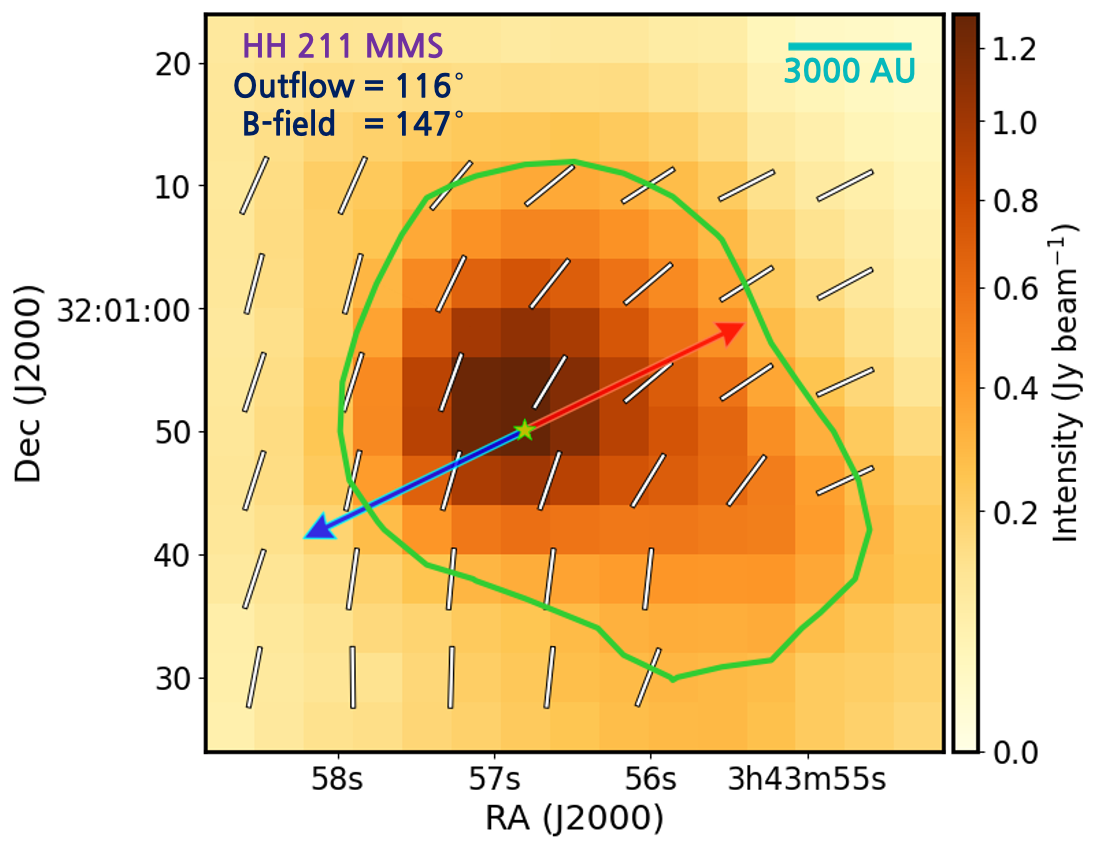}{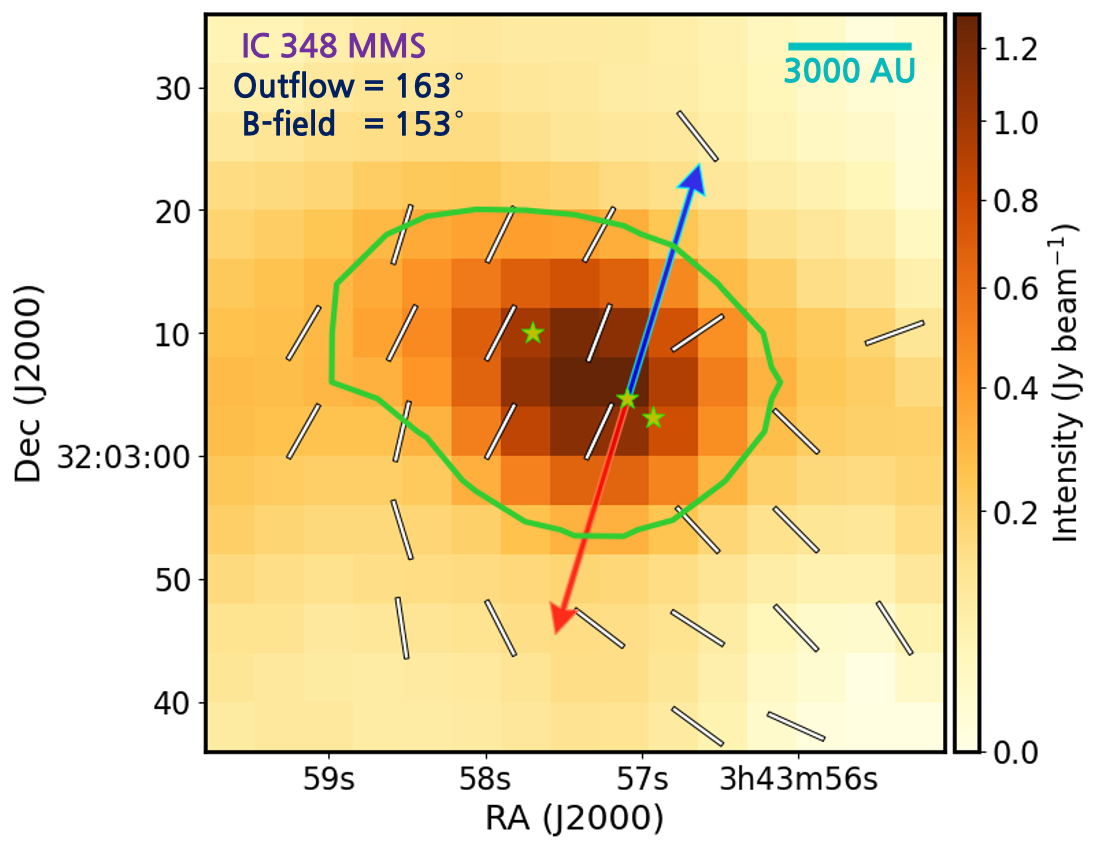}
\caption{The magnetic field orientations of HH 211 MMS (Left) and IC 348 MMS (Right). Green lines are at total intensity of 0.3 \jybm, indicative of the core regions. Red and blue lines represent red-shifted and blue-shifted outflow directions, respectively \citep{Rodr_guez_2014}, and stellar markers indicate the locations of protostars in the two cores \citep{Rodr_guez_2014}.
\label{fig:core}}
\end{figure*}

Note that the relative angles between magnetic fields and the filament here are based on projected directions. Real relative angles in the three-dimensional space could differ from the projected angles \citep{Doi_2020, Doi_2021}. Figure \ref{fig:cdf} shows the cumulative distribution function (CDF) of the projected angles between the magnetic fields and the filament. To show expected CDFs of projected angles between magnetic fields and filaments, we generated vectors in three-dimensional space, projected them into the plane of the sky, and constructed distributions of projected angles between two vectors, following the method given by \cite{Stephens_2017}. The observed CDF is qualitatively well fitted with the CDF when real three-dimensional angles are distributed between 45° and 80°. This result indicates that the magnetic field orientations in IC 348 are not random and are closer to a perpendicular configuration than a parallel one, even though the observed CDF does not perfectly match the CDF for three-dimensional angles ranging from 70° to 90°. It is important to note that the observed magnetic field orientations and the filament directions in IC 348 might result from a projection in a particular direction. In contrast, the expected CDFs are based on random projections. Detailed studies with modeling for three-dimensional structures of the magnetic fields and the filament are beyond the scope of this paper \citep[e.g.,][]{Tahani_2022,Tahani_2022_2}.

Figure \ref{fig:core} shows the magnetic field orientations of the two core regions sampled on an 8$''$ intervals. The mean field orientations are 147° for HH 211 MMS and 153° for IC 348 MMS within the defined core regions. The observed magnetic fields and outflow directions are relatively well aligned at the two core regions, even though previous statistical studies show that, in general, most outflows and magnetic field directions are randomly distributed \citep{Hull_2019,Doi_2020} or misaligned by 50° \citep{Yen_2021}.

\section{Polarization fraction versus Intensity} \label{sec:intensity and polarization fraction}

\begin{figure}[ht!]
\epsscale{1.17}
\plotone{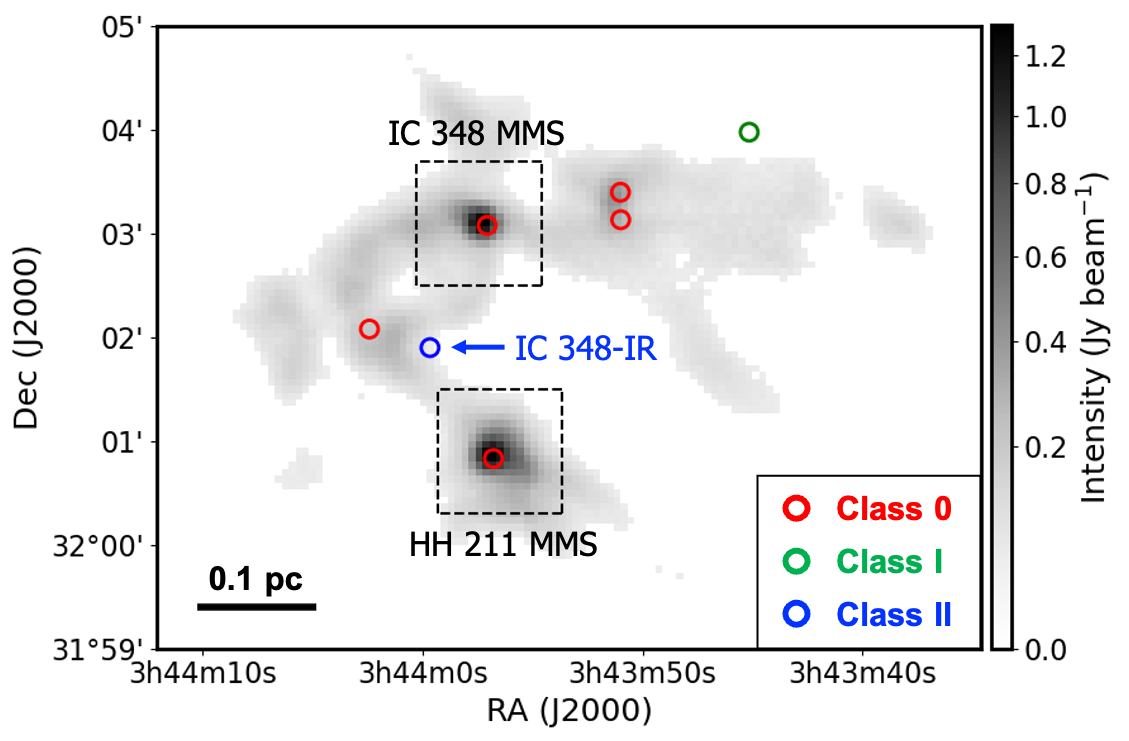}
\caption{Positions of young stellar objects in IC 348 from the VANDAM survey \citep{Tobin_2016}, are marked by circles: red, green, and blue ones for Class 0, I, and II sources. The black dashed squares show the area of HH 211 MMS and IC 348 MMS where the power-law and Ricean-mean fittings are carried out. A bright IR source IC 348-IR is located between HH 211 MMS and IC 348 MMS \citep{Strom_1974, McCaughrean_1994}. The background in grey is the 850 $\mu$m intensity.
\label{fig:PI_region}}
\end{figure}

The general trend of declining polarization efficiency with increased intensity in molecular clouds is well described by the power-law dependence of polarization fraction with intensity, $P \propto I^{-\alpha}$ \citep{Whittet_2008}. Perfect grain alignment results in $\alpha = 0$, while $\alpha = 1$ indicates that all polarization is due to the alignment of dust grains only on the surface of molecular clouds, and there is no polarization inside the clouds. If grain alignment decreases with increasing density, the power index $\alpha$ lies between 0 and 1 \citep{Whittet_2008,Jones_2015}.

We studied the power-law dependence of polarization fraction on intensity using both the single power-law model and the Ricean-mean method, as described by \cite{Pattle_2019}. The single power-law relation between polarization fraction and intensity is given by:

\begin{equation}
P_{\mathrm{debiased}} = p_{\sigma_{QU}}\left(\frac{I}{\sigma_{QU}}\right)^{-\alpha},
\label{eq:powerlaw}
\end{equation}
where $\sigma_{QU}$ is the rms noise in both the Stokes $Q$ and $U$ maps, $p_{\sigma_{QU}}$ represents the polarization fraction at the noise level of the data. If both the Stokes $Q$ and $U$ follow Gaussian distributions, non-debiased polarized intensity ($\sqrt{Q^2+U^2}$) is mathematically Rice-distributed. Rice distributions exhibit positive skewness with low S/N Stokes $Q$ and $U$ data, while approaching a Gaussian distribution in the high S/N limit. \cite{Pattle_2019} proposed the mean of the Rice distribution method to recover the $\alpha$ index, and we follow Equation (21) of \cite{Pattle_2019}. Their Monte Carlo simulations showed that the Ricean-mean method can accurately recover $\alpha$ values up to $\alpha \sim 0.6$. In contrast, the power-law method tends to overestimate $\alpha$ values since low S/N data points skew $\alpha$ toward 1 due to statistical noise. Even moderate S/N selection criteria tend to bias $\alpha$ values, as significant S/N is needed to distinguish between true power-law behavior and noise effects. For detailed descriptions, refer to \cite{Pattle_2019}.

We applied a power-law fit and Ricean-mean method to the two cores and the filament. The fitting areas of the two cores are shown as black dashed lines in Figure \ref{fig:PI_region}. For the filament, the whole region except the two black dashed areas is used for the fittings. In the power-law method, we used data points with $I/\delta I>10$, $PI/\delta PI>3$, and $\delta P<5\%$, whereas all data points within the fitting areas were used when applying the Ricean-mean method. We set $\sigma_{QU}$ to be the mean value of uncertainties in the Stokes $Q$ and $U$ maps, as suggested by \cite{Pattle_2019}. To find the values of $p_{\sigma_{QU}}$ and $\alpha$, we adopted a Markov Chain Monte Carlo (MCMC) sampling. Flat priors were used for both values, $\alpha$ was restricted to be between 0 and 1, and $p_{\sigma_{QU}}$ to be positive.

\begin{figure*}[ht!]
\epsscale{1.1}
\plottwo{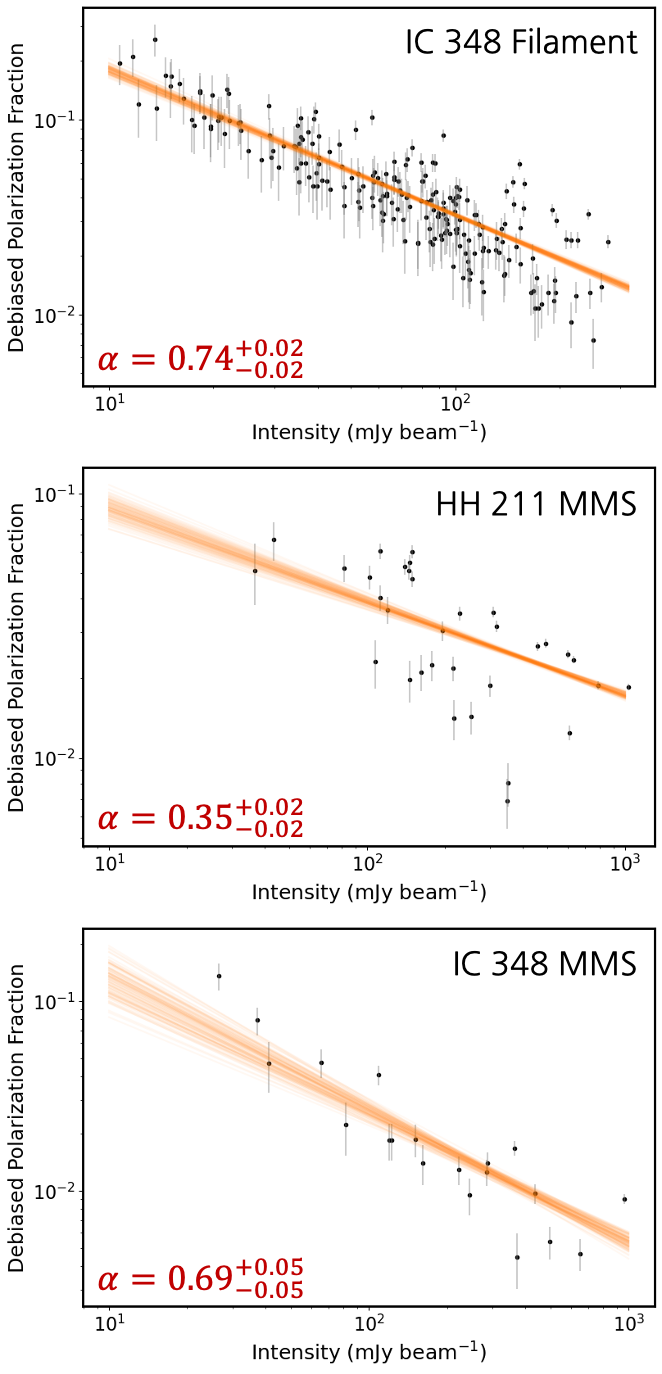}{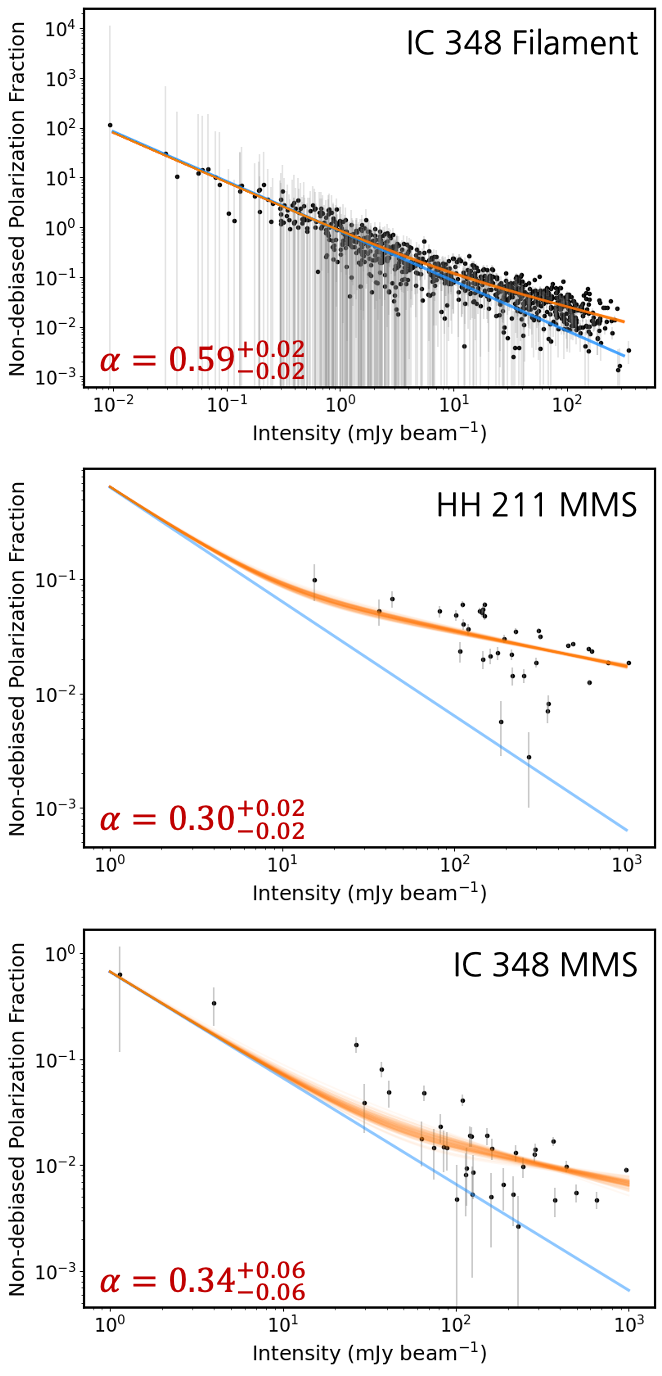}
\caption{Left: Debiased polarization fraction versus total intensity at 850 $\mu$m in the filament (top), HH 211 MMS (middle), and IC 348 MMS (bottom). Only data points with $I/\delta I>10$, $PI/\delta PI>3$, and  $\delta P<5\%$ are selected and are fitted with the power-law model. Orange lines represent 100 random draws from the posterior distribution. Right: Non-debiased polarization fraction versus total intensity at 850 $\mu$m in the filament (top), HH 211 MMS (middle), and IC 348 MMS (bottom). All data points within the defined areas are selected and are fitted with the Ricean-mean model. Orange lines represent 100 random draws from the posterior distribution, while the blue lines indicate the case of $\alpha = 1$. The best-fit alpha values are indicated at the bottom left of each panel.
\label{fig:PI_relation}}
\end{figure*}

Figure \ref{fig:PI_relation} shows the best-fit results from the power-law and the mean of the Rice distribution methods, and the recovered $p_{\sigma_{QU}}$ and $\alpha$ values are summarized in Table \ref{tab:PI}. The uncertainties of individual parameters are based on the 16th and 84th percentiles of the sample distributions. In all cases, the Ricean-mean method gives smaller $\alpha$ values as expected by \cite{Pattle_2019}, which supports the argument that the power-law method could overestimate $\alpha$ values in comparison with the Ricean-mean method. Therefore, we adopt the results from the Ricean-mean method.

The power indices obtained by the Ricean-mean method are 0.59, 0.30, and 0.34 for the filament, HH 211 MMS and IC 348 MMS, respectively. These results indicate that a significant fraction of dust grains should be aligned with the magnetic field even in the high-density regions, justifying the use of polarization observations to infer the magnetic field properties of the area. Furthermore, it is noticeable that the power indices of the two cores are smaller than that of the filament. This result implies that some mechanisms effectively align dust grains in high-density regions of the cores.

\begin{deluxetable*}{ccccc}
\tablenum{1}
\tablecaption{Fitted $p_{\sigma_{QU}}$ and $\alpha$ values in $P$ versus $I$ relations \label{tab:PI}}
\tablewidth{0pt}
\tablehead{\colhead{Method} & \colhead{Description} & \colhead{Filament} & \colhead{HH 211 MMS} & \colhead{IC 348 MMS}
}
\startdata 
Power-law & $p_{\sigma_{QU}}$ & $1.50^{+0.16}_{-0.14}$ & $0.26^{+0.03}_{-0.03}$ & $0.98^{+0.35}_{-0.26}$ \\ [6pt]
          & $\alpha$ & $0.74^{+0.02}_{-0.02}$ & $0.35^{+0.02}_{-0.02}$ & $0.69^{+0.05}_{-0.05}$  \\ [6pt]
%\hline
Mean of the Rice distribution & $p_{\sigma_{QU}}$ & $0.48^{+0.04}_{-0.04}$ & $0.17^{+0.02}_{-0.02}$ & $0.09^{+0.04}_{-0.03}$ \\ [6pt]
                              & $\alpha$ & $0.59^{+0.02}_{-0.02}$ & $0.30^{+0.02}_{-0.02}$ & $0.34^{+0.06}_{-0.06}$  \\
\enddata
\label{tab}
\end{deluxetable*}

\section{Magnetic field strength} \label{sec:magnetic field strength}

In this section, we estimate the magnetic field strength in the IC 348 star-forming region using the Davis-Chandrasekhar-Fermi (DCF) \citep{Davis_1951_2,Chand_1953} method and its alternative version for compressed medium \citep[ST method;][]{Skalidis_2021}. The DCF method assumes that turbulence collectively shakes the aligned dust grains, thereby distorting the polarization direction. If magnetic fields are strong enough, the magnetic fields are little distorted, while if magnetic fields are weak, they are disturbed significantly. Therefore, it is possible to estimate the magnetic field strength by measuring the degree of dispersion in magnetic field orientations.

There have been numerous efforts to modify and refine the DCF method to estimate the  magnetic field strength accurately \citep[e.g.,][]{Hildebrand_2009,Houde_2009,Cho_2016,Skalidis_2021}{}. One example is the ST method \citep{Skalidis_2021}. \cite{Skalidis_2021} pointed out that DCF assumes incompressible fluids and ignores the compressible modes. Since the interstellar medium is highly compressible, the DCF assumption may not be generally applicable and lead to an inaccurate estimation of field strengths. Thus, they proposed an alternative method containing the presence of compressible turbulence. However, it is claimed that the ST method also contains uncertainties and interpretation of this method should also be done with caution \citep{Li_2022,Lazarian_2022,Liu_2022}. Detailed studies with the physical interpretation of the DCF method and its various modified versions are beyond the scope of this paper \citep[e.g.,][]{Lazarian_2022,Liu_2022}{}. In this paper, we applied both DCF and ST methods to provide a range of  results from them.

The equations of DCF and ST methods are as follows:

\begin{equation}
B_{\mathrm{pos}} = f \hspace{0.1cm} \sqrt{4 \pi \rho} \hspace{0.1cm} \frac{\delta v}{\delta \theta} \quad (\mathrm{DCF}),
\label{eq:dcf1}
\end{equation}

\begin{equation}
B_{\mathrm{pos}} = \sqrt{4 \pi \rho} \hspace{0.1cm} \frac{\delta v}{\sqrt{2 \delta \theta}} \quad (\mathrm{ST}),
\label{eq:st1}
\end{equation}
where $\rho$ is the gas mass density, $\delta v$ is the velocity dispersion, and $\delta \theta $ is the dispersion in polarization angles representing magnetic field orientations. 

DCF has a correcting factor $f$, which is commonly used for modifying the original DCF method. \cite{Lazarian_2022} pointed out that the correcting factor depends on the properties of the turbulence. \cite{Liu_2022} also argued that the correcting factor decreases as the density of the region increases based on the results of three independent simulations \citep{Ostriker_2001,Padoan_2001,Liu_2021}. In this paper, we applied $f = 0.5$, which is commonly used.

To apply the DCF and ST methods to our target, we converted Equations \ref{eq:dcf1} and \ref{eq:st1} into the formulations given by \cite{Crutcher_2004} as follows:

\begin{equation}
B_{\mathrm{pos}} \hspace{0.1cm} (\mu \mathrm{G}) \approx 9.42 \hspace{0.1cm} \sqrt{n(\mathrm{H}_2)\hspace{0.1cm}(\mathrm{cm}^{-3})} \hspace{0.1cm} \frac{\Delta V\hspace{0.1cm}(\mathrm{km} \hspace{0.05cm} \mathrm{s}^{-1})}{\sigma_{\theta}\hspace{0.1cm}(\mathrm{deg})} \quad (\mathrm{DCF}),
\label{eq:dcf}
\end{equation}

\begin{equation}
B_{\mathrm{pos}} \hspace{0.1cm} (\mu \mathrm{G}) \approx 1.76 \hspace{0.1cm} \sqrt{n(\mathrm{H}_2)\hspace{0.1cm}(\mathrm{cm}^{-3})} \hspace{0.1cm} \frac{\Delta V\hspace{0.1cm}(\mathrm{km} \hspace{0.05cm} \mathrm{s}^{-1})}{\sqrt{\sigma_{\theta}\hspace{0.1cm}(\mathrm{deg})}} \quad (\mathrm{ST}),
\label{eq:st}
\end{equation}
where $n(\mathrm{H}_2)$ is the number density of hydrogen molecules, $\Delta V$ is the FWHM velocity dispersion, $\sigma_{\theta}$ is the angular dispersion in polarization angles and $B_{\mathrm{pos}}$ is the plane-of-sky magnetic field strength. We adopt $f = 0.5$ in the DCF analysis. Using Equations \ref{eq:dcf} and \ref{eq:st}, we measured the magnetic field strengths of the two core regions and the filament defined in Figure \ref{fig:positions}, and studied the roles of magnetic fields in both core ($\sim$0.05 pc) and filament scales ($\sim$0.5 pc). The results are summarized in Table \ref{tab:B}.

\subsection{Angular Dispersion of Polarization} \label{subsec:angular dispersion in IC 348}

We constructed the observed polarization angle map using the  Stokes $Q$ and $U$ maps with Equation \ref{eq:theta}. Then, to make a background large-scale polarization map, we applied the unsharp masking method \citep{Pattle_2017} which smoothes the $Q$ and $U$ maps with the 3 $\times$ 3 boxcar kernel. Since a larger boxcar filter could overly smooth the non-turbulent magnetic field curvature of the filament, we chose a 3 $\times$ 3 boxcar filter, corresponding to a length of 0.05 pc, which is comparable to a half of the typical filament's width \citep{Andr__2014}. Subtracting the background polarization map from the original polarization angle map provides residual polarization angles:
\begin{equation}
\Delta \theta = \theta_{\mathrm{data}} - \theta_{\mathrm{bg}},
\label{eq:ang_disp}
\end{equation}
where $\theta_{\mathrm{data}}$ and $\theta_{\mathrm{bg}}$ represent the observed and smoothed polarization angles, respectively, while $\Delta \theta$ indicates the residual angles.
To calculate the angular dispersion of the three different regions defined above, we cropped the residual map into individual areas and estimated the dispersion from each cropped residual angle map. In this process, the polarization angles in core regions are sampled on an 8$''$ grid.

\begin{figure*}[ht!]
\gridline{\fig{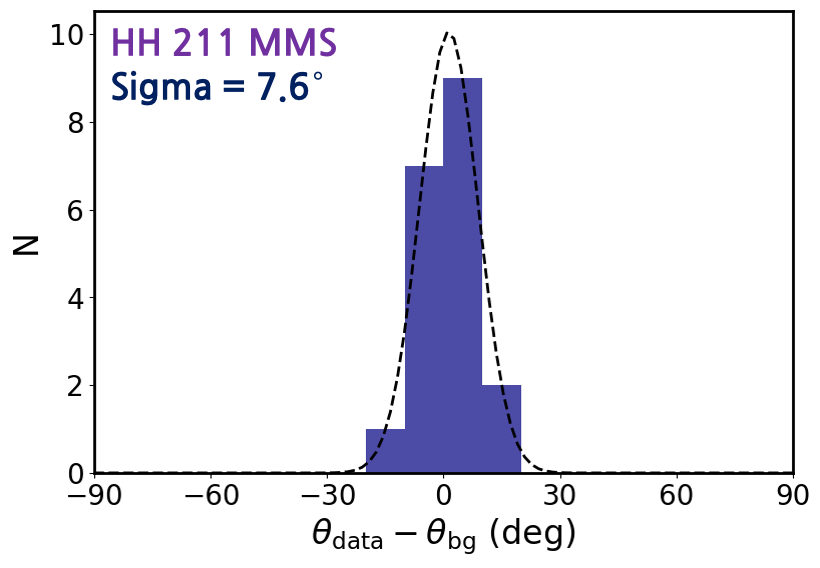}{0.3\textwidth}{(a)}
          \fig{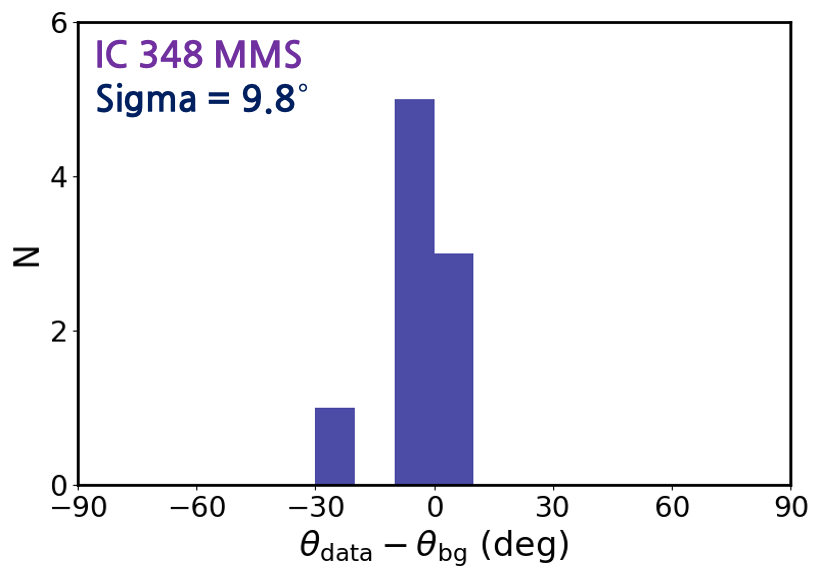}{0.3\textwidth}{(b)}
          \fig{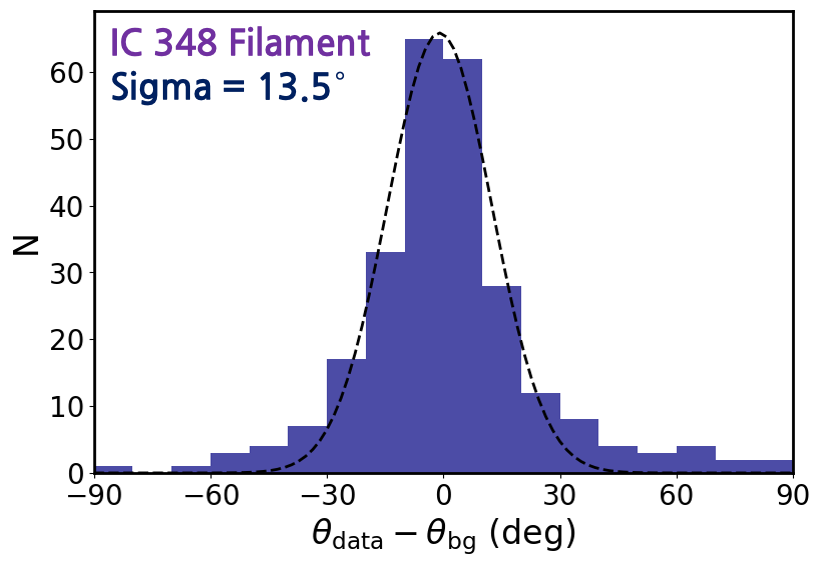}{0.3\textwidth}{(c)}
          }
\caption{Residual polarization angle ($\Delta \theta$) distributions of (a) HH 211 MMS, (b) IC 348 MMS, and (c) the filament. The dashed lines are the fitted Gaussian functions, whose sigmas are shown in each panel. For IC 348 MMS, the standard deviation of the distribution is adopted as the angular dispersion of the region.
\label{fig:ang_disp}}
\end{figure*}

Figure \ref{fig:ang_disp} shows the distributions of $\Delta \theta$ in HH 211 MMS, IC 348 MMS, and the filament. We fitted a Gaussian function to each distribution and adopted its standard deviation as the angular dispersion of each area. For IC 348 MMS, instead of the Gaussian fitting, the standard deviation of the distribution is adopted due to the lack of data points. The angular dispersions of HH 211 MMS, IC 348 MMS, and the filament are 7.6°$\pm$0.1°, 9.8°, and 13.5°$\pm$ 0.7°, respectively.

\subsection{Number Density} \label{subsec:number density of IC 348}

Number densities are determined from 450 $\mu$m and 850 $\mu$m intensity maps observed by SCUBA-2. We applied a Gaussian convolution to the 450 $\mu$m intensity map to match the angular resolution to the 850 $\mu$m intensity map. We assumed that dust continuum in IC 348 is optically thin at both wavelengths. The observed intensity in optically thin cases is given by:
\begin{eqnarray}
I_{\nu} & = & \tau_{\nu}B_{\nu}(T) \nonumber \\
& = & \kappa(\nu)\Sigma_{\mathrm{d}}B_{\nu}(T) \nonumber \\
& = & \mu m_{\mathrm{H}}\kappa(\nu)N(\mathrm{H}_2)B_{\nu}(T),
\label{eq:intens}
\end{eqnarray}
where $I_{\nu}$ is the intensity at a frequency $\nu$, $\tau_{\nu}$ is the optical depth, $\Sigma_{\mathrm{d}}$ is the mass column density, $\mu$ is the mean molecular weight of the molecular cloud, $m_{\mathrm{H}}$ is the mass of a hydrogen atom, $\kappa(\nu)$ is the dust mass absorption coefficient at a frequency $\nu$ \citep{Hildebrand_1983}, $N(\mathrm{H}_2)$ is the column density of molecular hydrogen, and $B_{\nu}(T)$ is the blackbody radiation at a dust temperature $T$. We adopted $\mu = 2.86$, assuming the proportion of hydrogen in gas mass is 70\% \citep{Kirk_2013}.

The dust opacity function is as follows:
\begin{equation}
\kappa(\nu) = \kappa_{\nu_0}\left(\frac{\nu}{\nu_0}\right)^{\beta},
\label{eq:opacity}
\end{equation}
where $\kappa_{\nu_0}$ is the dust opacity at a frequency $\nu_{0}$ and $\beta$ is a dust opacity spectral index. We take a $\beta$ value of 1.7 \citep[e.g.,][]{Lin_2016}, and $\kappa_{\nu_0} = 0.1$ cm$^2$ g$^{-1}$ at $\nu_0 = 1$ THz, assuming a gas-to-dust mass ratio of 100 \citep{Beckwith_1991}.

To measure the column density and dust temperature of the target region, we follow the method given by \cite{Pattle_2017_2}. We divided the 850 $\mu$m intensity map by the 450 $\mu$m intensity map using Equation \ref{eq:intens}. Then, the column density term is eliminated, leaving only the temperature-related equation as:
\begin{equation}
\frac{I_{850}}{I_{450}} = \left(\frac{\nu_{850}}{\nu_{450}}\right)^{3+\beta}  \left(\frac{e^{\frac{h\nu_{450}}{k_{B}T}}-1}{e^{\frac{h\nu_{850}}{k_{B}T}}-1}\right).
\label{eq:temp}
\end{equation}

\begin{figure*}[ht!]
\epsscale{1.15}
\plotone{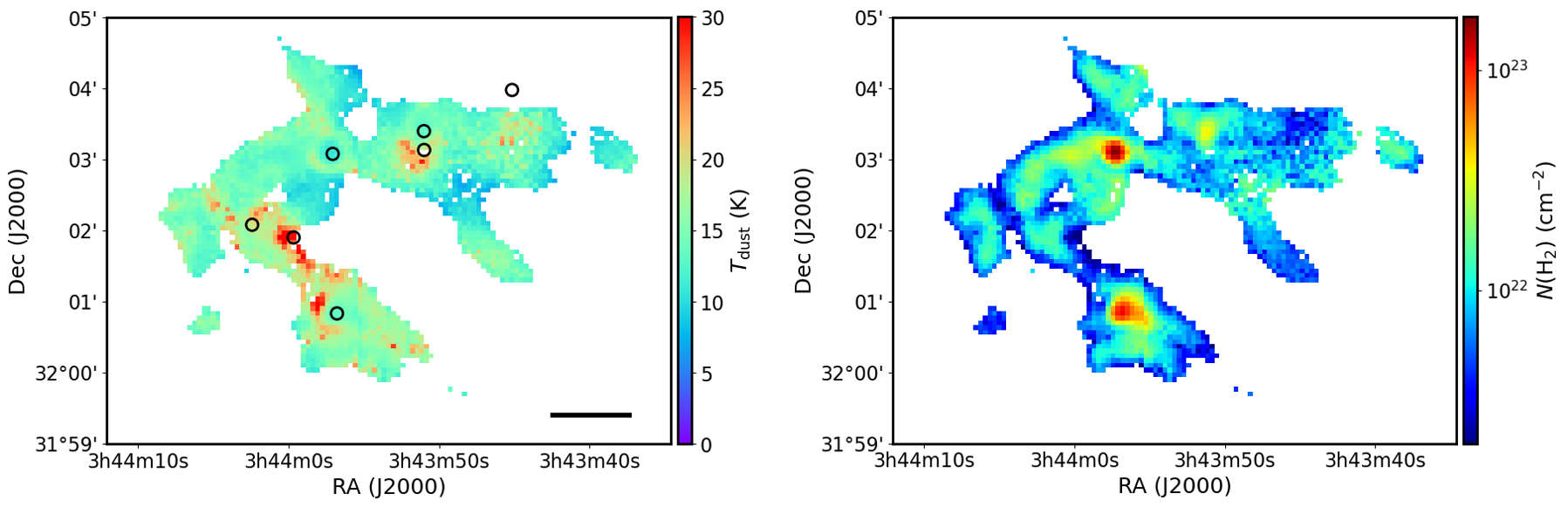}
\caption{The estimated dust temperature (left) and column density (right) maps of the IC 348 star-forming region. The black circles in the left panel indicate the positions of young stellar objects, as in Figure \ref{fig:PI_region}. 
\label{fig:col_den}}
\end{figure*}

We solved Equation \ref{eq:temp} to measure the dust temperature. Then, we inferred the column density of each pixel using Equation \ref{eq:intens}. Figure \ref{fig:col_den} shows the estimated dust temperature and column density of IC 348. Dust temperature is high near the young stellar objects, especially near IC 348-IR marked in Figure \ref{fig:PI_region}. Two core regions have column densities greater than $10^{23}$ cm$^{-2}$ at their central areas, and the filamentary structure has a column density scale of $10^{22}$ cm$^{-2}$.

The column densities of HH 211 MMS, IC 348 MMS, and the filament are obtained by averaging the pixel values of the defined areas. We calculated the uncertainty in the inferred column density by applying error propagation through Equations \ref{eq:intens}-\ref{eq:temp}. We assumed the uncertainty on the reference dust opacity $\kappa_{\nu 0}$ is 50\% \citep{Roy_2014}, and the uncertainty of the dust opacity spectral index $\beta$ is $\pm$ 0.3 \citep{Sadavoy_2013,Sadavoy_2016}. Uncertainties in 450 $\mu$m and 850 $\mu$m intensities are about 15$\%$ and 6$\%$ respectively \citep{Mairs_2021}.  When considering these uncertainties, the systematic uncertainty in calculated column densities is about 67$\%$. Consequently, the measured column densities of each area are as follows: $N(\mathrm{H}_2) = (4.1 \pm 2.9) \times 10^{22}$ cm$^{-2}$, $ (6.2 \pm 4.1) \times 10^{22}$ cm$^{-2}$, and $(9.2 \pm 6.2) \times 10^{21}$ cm$^{-2}$ for HH 211 MMS, IC 348 MMS, and the filament, respectively.

For calculating number densities of the region, we assumed that the line-of-sight depth of the cloud is equal to the width of the filament. We calculated the width of the filament, which is 0.050 pc, by dividing the defined filament area (Figure \ref{fig:positions}) by the length of the filament skeleton in the left panel of Figure \ref{fig:fil_direc}. The number density $n(\mathrm{H}_2)$ is estimated through
\begin{equation}
n(\mathrm{H}_2) = \frac{N(\mathrm{H}_2)}{W},
\label{eq:num_den2}
\end{equation}
where $N(\mathrm{H}_2)$ is the column density, and $W$ is the width or line-of-sight depth of the region. Assuming the uncertainty of the width is 50\%, the measured number densities of HH 211 MMS, IC 348 MMS, and the filament are  $n(\mathrm{H}_2) = (2.7 \pm 2.3) \times 10^{5}$ cm$^{-3}$, $ (4.0 \pm 3.3) \times 10^{5}$ cm$^{-3}$, and $ (6.0 \pm 5.1) \times 10^{4}$ cm$^{-3}$, respectively.

\subsection{Velocity Dispersion} \label{subsec:velocity dispersion in IC 348}

\begin{figure*}[ht!]
\epsscale{1.15}
\plotone{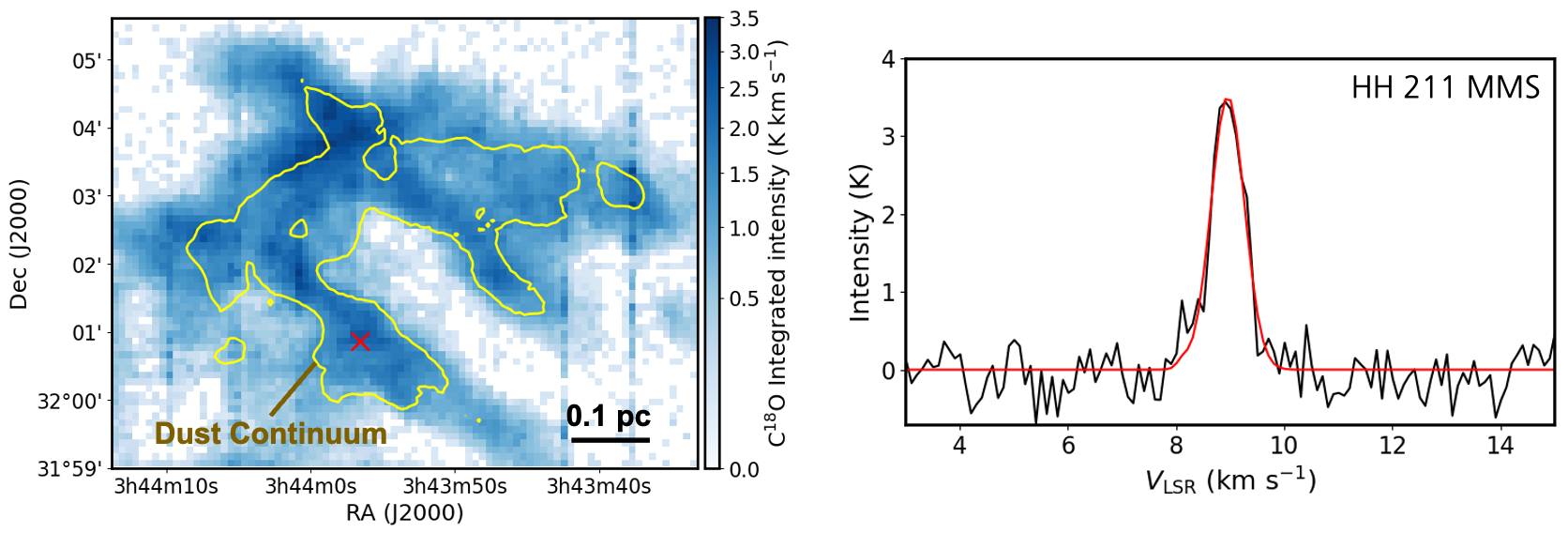}
\caption{Left: The integrated intensity map of the C$^{18}$O $J = 3 \rightarrow 2$ emissions. Yellow contour indicates the 850 $\mu$m dust continuum at 0.05 Jy beam$^{-1}$ level. The red cross mark shows the position of the spectrum shown in the right panel. Right: An example C$^{18}$O spectral line toward HH 211 MMS. The observed spectrum and its Gaussian fitting result are shown as the black and red lines, respectively.
\label{fig:vel_disp}}
\end{figure*}

Average velocity dispersions in IC 348 are calculated from the C$^{18}$O $J = 3 \rightarrow 2$ observations made using the the Heterodyne Array Receiver Program (HARP: \cite{Buckle_2009}). The left panel of Figure \ref{fig:vel_disp} shows that C$^{18}$O emission is spatially well correlated with the dust continuum. To measure C$^{18}$O velocity dispersion, we fitted a Gaussian profile to a C$^{18}$O spectrum of each pixel and accepted the Gaussian width as the C$^{18}$O velocity dispersion of the pixel. We measured the average velocity dispersion over the pixels within the defined areas to calculate the C$^{18}$O velocity dispersion of the two cores and the filament. Measured C$^{18}$O velocity dispersions for HH 211 MMS, IC 348 MMS, and the filament are $\sigma_{\nu,\mathrm{C}^{18}\mathrm{O}} = 0.27 \pm 0.04$ km s$^{-1}$, $0.30 \pm 0.04$ km s$^{-1}$, and $0.26 \pm 0.07$ km s$^{-1}$, respectively.

From the C$^{18}$O velocity dispersion, we estimated the non-thermal velocity dispersion by removing the thermal component:
\begin{equation}
\sigma^2_{v} = \sigma^2_{v, \mathrm{C}^{18}\mathrm{O}} - \frac{k_{\mathrm{B}} T}{m_{\mathrm{C}^{18}\mathrm{O}}},
\label{eq:vel_disp}
\end{equation}
where $\sigma_{v}$ is the non-thermal velocity dispersion, $\sigma_{v, \mathrm{C}^{18}\mathrm{O}}$ is the total velocity dispersion of C$^{18}$O, $m_{\mathrm{C}^{18}\mathrm{O}} = 30$ amu (the mass of the C$^{18}$O molecule), $k_{\mathrm{B}}$ is the Boltzmann constant and $T$ is the gas temperature. Assuming that the gas temperature is the same as the dust temperature, the gas temperature values for the thermal motion were obtained by averaging the estimated dust temperatures in the defined areas. They are $16.2\pm6.0$ K and $13.4\pm3.9$ K at HH 211 MMS and IC 348 MMS, and $14.8\pm4.9$ K at the filament.

Finally, we derived non-thermal velocity dispersions of $\sigma_{v}=0.26 \pm 0.04$ km s$^{-1}$, $0.29 \pm 0.04$ km s$^{-1}$, and $0.25 \pm 0.07$ km s$^{-1}$ for HH 211 MMS, IC 348 MMS and the filament, respectively. These correspond to velocity FWHMs of $\Delta V=0.62 \pm 0.09$ km s$^{-1}$, $0.68 \pm 0.10$ km s$^{-1}$, and $0.60 \pm 0.18$ km s$^{-1}$.

\subsection{Magnetic Field Strength} \label{subsec:magnetic field strength of IC 348}

Using Equations \ref{eq:dcf} and \ref{eq:st}, we determined magnetic field strengths of the two cores and the filament. Table \ref{tab:B} lists the field strengths from the DCF and ST methods. The results from the DCF analysis are $B_{\mathrm{pos}} =$ $394 \pm 170$ $\mu$G, $417 \pm 182$ $\mu$G, and $102 \pm 52$ $\mu$G for HH 211 MMS, IC 348 MMS, and the filament, respectively. The respective magnetic field strengths using the ST method are $B_{\mathrm{pos}} =$ $203 \pm 87$ $\mu$G, $244 \pm 106$ $\mu$G, and $ 70 \pm 35$ $\mu$G.

The mass-to-flux ratio $\lambda$ is defined in units of the critical value which is inferred using the equation below, given by \cite{Crutcher_2004}:
\begin{equation}
\lambda = 7.6 \times 10^{-21} \hspace{0.1cm}\frac{N(\mathrm{H}_2)\hspace{0.1cm}(\mathrm{cm}^{-2})}{B_{\mathrm{pos}}\hspace{0.1cm}(\mu \mathrm{G})}.
\label{eq:crit}
\end{equation}
This parameter is used to measure the relative importance between gravity and magnetic field in molecular clouds \citep{Nakano_1978}. When $\lambda > 1$ (magnetically supercritical), gravity dominates over the magnetic field. In contrast, $\lambda < 1$ (magnetically subcritical) indicates that the magnetic field is strong enough to prevent gravitational collapse.

The mass-to-flux ratios based on the magnetic field strengths of both the DCF and ST methods are given in Table \ref{tab:B}. For the DCF method, we find values of  $\lambda=0.79 \pm 0.34$, $1.12 \pm 0.49$, and $0.68 \pm 0.35$ for HH 211 MMS, IC 348 MMS, and the filament, respectively. With the ST method, the respective ratios are $\lambda=1.54 \pm 0.66$, $1.92 \pm 0.84$ and $1.00 \pm 0.50$. The core regions have larger $\lambda$ values than the filament. These results suggest that the cores are likely to be magnetically supercritical or in balance, while the filament is likely to be magnetically subcritical. Note that the observed mass-to-flux ratios may overestimate the true values since only the plane-of-sky magnetic field components are considered \citep{Crutcher_2004}. Here, we did not apply geometric corrections to the observed mass-to-flux ratios due to lack of line-of-sight magnetic field information.

\begin{deluxetable*}{ccccc}
\tablenum{2}
\tablecaption{The measured properties for estimating the magnetic field strengths in the IC 348 star-forming region \label{tab:B}}
\tablewidth{0pt}
\tablehead{\colhead{Parameter} & \colhead{Description} & \colhead{HH 211 MMS} & \colhead{IC 348 MMS} & \colhead{Filament}
}
\startdata
$\sigma_{\theta}$ & Angular dispersion & 7.6°$\pm$0.1° & 9.8° & 13.5°$\pm$0.7°\\
$N(\mathrm{H}_2)$ & Hydrogen column density & $(4.1 \pm 2.9) \times 10^{22}$ cm$^{-2}$ & $(6.2 \pm 4.1) \times 10^{22}$ cm$^{-2}$ & $(9.2 \pm 6.2) \times 10^{21}$ cm$^{-2}$ \\
$n(\mathrm{H}_2)$ & Hydrogen number density & $(2.7 \pm 2.3) \times 10^{5}$ cm$^{-3}$ & $(4.0 \pm 3.3) \times 10^{5}$ cm$^{-3}$ & $(6.0 \pm 5.1) \times 10^{5}$ cm$^{-3}$ \\
$T_{\mathrm{dust}}$ & Dust temperature & $16.2 \pm 6.0$ K & $13.4 \pm 3.9$ K & $14.8 \pm 4.9$ K \\
$M$ & Mass & $2.45 \pm 1.57$ $M_{\odot}$ & $2.09 \pm 1.25$ $M_{\odot}$ & $16.09 \pm 9.95$ $M_{\odot}$ \\
$R$, $W$ & Radius, Width & $0.029$ pc & $0.022$ pc & $0.050$ pc \\
$\sigma_{v}$ & Non-thermal velocity dispersion & $0.26 \pm 0.04$ km s$^{-1}$ & $0.29 \pm 0.04$ km s$^{-1}$ & $0.25 \pm 0.07$ km s$^{-1}$ \\
$\Delta V$ & Velocity FWHM & $0.62 \pm 0.09$ km s$^{-1}$ & $0.68 \pm 0.10$ km s$^{-1}$ & $0.60 \pm 0.18$ km s$^{-1}$ \\
$B_{\mathrm{pos},\mathrm{DCF}}$ & Magnetic field strength (DCF) & $394 \pm 170$ $\mu$G & $417 \pm 182$ $\mu$G & $102 \pm 52$ $\mu$G \\
$\lambda_{\mathrm{DCF}}$ & Mass-to-flux ratio (DCF) & $0.79 \pm 0.34$ & $1.12 \pm 0.49$ & $0.68 \pm 0.35$  \\
$B_{\mathrm{pos},\mathrm{ST}}$ & Magnetic field strength (ST) & $203 \pm 87$ $\mu$G & $244 \pm 106$ $\mu$G & $70 \pm 35$ $\mu$G \\
$\lambda_{\mathrm{ST}}$ & Mass-to-flux ratio (ST) & $1.54 \pm 0.66$ & $1.92 \pm 0.84$ & $1.00 \pm 0.50$  \\
\enddata
\end{deluxetable*}

\section{Energy balance in the core regions} \label{sec:energy calculations}

The interplay between gravity, thermal pressure, turbulence, and magnetic fields plays a role in the star formation process. Therefore, it is vital to probe the energy balance of molecular clouds at different scales to understand the interaction between various factors. In this section, we measure (1) magnetic energies, (2) gravitational energies, (3) thermal energies, and (4) turbulent energies in HH 211 MMS and IC 348 MMS and compare the relative importance between them. The results are summarized in Table \ref{tab:E}.

The magnetic energy is calculated as
\begin{equation}
E_{B} = \frac{B^{2}}{8\pi} \hspace{0.1cm} V
\label{eq:E_B}
\end{equation}
in cgs units, where $V$ is the region's volume, and $B$ is the magnetic field strength. We assumed that the cores are a sphere whose radius to be that of a circle with the same area to the core. The measured radii are 0.029 pc and 0.022 pc for HH 211 MMS and IC 348 MMS, respectively. When adopting the DCF magnetic field strengths, the inferred magnetic energies are $E_{B} = (1.8 \pm 1.6) \times 10^{43}$ erg in HH 211 MMS and $(8.8 \pm 7.7) \times 10^{42}$ erg in IC 348 MMS. The ST field strengths result in $E_{B} = (4.8 \pm 4.1) \times 10^{42}$ erg for HH 211 MMS and $(3.0 \pm 2.6) \times 10^{42}$ erg for IC 348 MMS.

The gravitational energy of an uniform sphere is given by
\begin{equation}
E_{G} = -\frac{3}{5} \frac{GM^{2}}{R},
\label{eq:E_G}
\end{equation}
where $R$ is the radius and $M$ is the mass of the sphere. Mass can be estimated from the dust thermal emission by the following equation \citep[e.g.,][]{Hildebrand_1983,Kwon_2009}:
\begin{equation}
M = \frac{F_\nu D^2}{\kappa_{\nu}B_{\nu}(T)},
\label{eq:mass}
\end{equation}
where $F_{\nu}$ is the total flux, $D$ is the distance, $\kappa_{\nu}$ is the opacity and $B_{\nu}$ is blackbody radiation. The masses are estimated as $(2.45 \pm 1.57)$ $\mathrm{M}_{\odot}$ for HH 211 MMS and $(2.09 \pm 1.25)$ $\mathrm{M}_{\odot}$ for IC 348 MMS. These correspond to gravitational energies of $E_{G} = (-1.1 \pm 1.4) \times 10^{43}$ erg and $(-1.0 \pm 1.2) \times 10^{43}$ erg for HH 211 MMS and IC 348 MMS, respectively. Note that the estimated gravitational energies from Equation \ref{eq:E_G} are lower limit (in absolute value), since the mass is expected to be concentrated near the center of the cores.

\begin{deluxetable*}{cccc}
\tablenum{3}
\tablecaption{\label{tab:E} Energy distributions in the IC 348 core regions}
\tablewidth{0pt}
\tablehead{\colhead{Parameter} & \colhead{Description} & \colhead{HH 211 MMS} & \colhead{IC 348 MMS}
}
\startdata
$E_{G}$ & Gravitational energy & $-11 \pm 14$ & $-10 \pm 12$\\
$E_{B,\mathrm{DCF}}$ & Magnetic energy (DCF) & $18 \pm 16$ & $8.8 \pm 7.7$\\
$E_{B,\mathrm{ST}}$ & Magnetic energy (ST) & $4.8 \pm 4.1$ & $3.0 \pm 2.6$\\
$E_{\mathrm{thermal}}$ & Thermal energy & $3.4 \pm 2.5$ & $2.4 \pm 1.6$ \\
$E_{\mathrm{non}\mbox{-}\mathrm{thermal}}$ & Turbulent energy & $4.9 \pm 3.5$ & $5.2 \pm 3.5$ \\
\enddata
\tablecomments{Unit: $10^{42}$ erg}
\end{deluxetable*}

The thermal energy is calculated as
\begin{equation}
E_{\mathrm{thermal}} = \frac{3}{2} M c_{s}^2, 
\label{eq:E_thermal}
\end{equation}
where $c_{s}$ is the sound speed in the gas:

\begin{equation}
c_{s} = \sqrt{\frac{k_{B}T}{\mu m_{\mathrm{H}}}}.
\label{eq:sound}
\end{equation}
The gas temperature is assumed to be the same as the dust temperature. Using the estimated masses and dust temperatures of the core regions, thermal energies are calculated to be $E_{\mathrm{thermal}} = (3.4 \pm 2.5) \times 10^{42}$ erg for HH 211 and $(2.4 \pm 1.6) \times 10^{42}$ erg for IC 348 MMS.

The turbulent energy or non-thermal kinetic energy is calculated as
\begin{equation}
E_{\mathrm{non}\mbox{-}\mathrm{thermal}} = \frac{3}{2} M \sigma_{\nu}^2,
\label{eq:E_nonthermal}
\end{equation}
where $\sigma_{\nu}$ is a non-thermal velocity dispersion of the gas. 
The inferred masses and non-thermal velocity dispersions of the core regions give turbulent energies of $E_{\mathrm{non}\mbox{-}\mathrm{thermal}} = (4.9 \pm 3.5) \times 10^{42}$ erg and $(5.2 \pm 3.5) \times 10^{42}$ erg for HH 211 MMS and IC 348 MMS, respectively. 

Regarding HH 211 MMS, the magnetic energy based on the DCF result dominates the energy balance. However, in the ST method gravity dominates the system, and magnetic, thermal, and turbulent energies are comparable to one another. For IC 348 MMS, the gravity and magnetic field dominate the energy balance based on the DCF result. In contrast, with the ST method, gravity dominates the system, and the turbulent energy is larger than magnetic and thermal energies.

\section{Discussion} \label{sec:discussion}
\subsection{Grain Alignment in the Filament and the Two Cores} \label{subsec:alignment size}
 
The slopes ($\alpha$ index) between polarization fraction and intensity reported in Section \ref{sec:intensity and polarization fraction} and Figure \ref{fig:PI_relation} are shallower than 1, which suggests that there is a working alignment mechanism even in dense regions. Regarding the filament first, according to the Ricean-mean fitting that is presumably more reliable than the power-law fitting, the slope is 0.59. This could be understood with grain alignment enhanced by the radiation from the young stellar objects along the filament as shown in the positions (Figure \ref{fig:PI_region}) and the temperature (the left panel of Figure \ref{fig:col_den}) \citep[e.g.,][]{Lazarian_2007}. \cite{Pattle_2019} suggested that a smaller $\alpha$ value in the Oph A region ($\alpha = 0.34$) is mainly due to the nearby two B stars of the Oph A region. \cite{Lyo_2021} proposed that the lower $\alpha$ value for NGC 2071IR ($\alpha = 0.36$) is attributed to the additional radiation from the central three infrared young stellar objects. In IC 348, radiations from young stellar objects along the filament could cause the shallower slope in the filament.

\begin{figure}[t!]
\epsscale{1.17}
\plotone{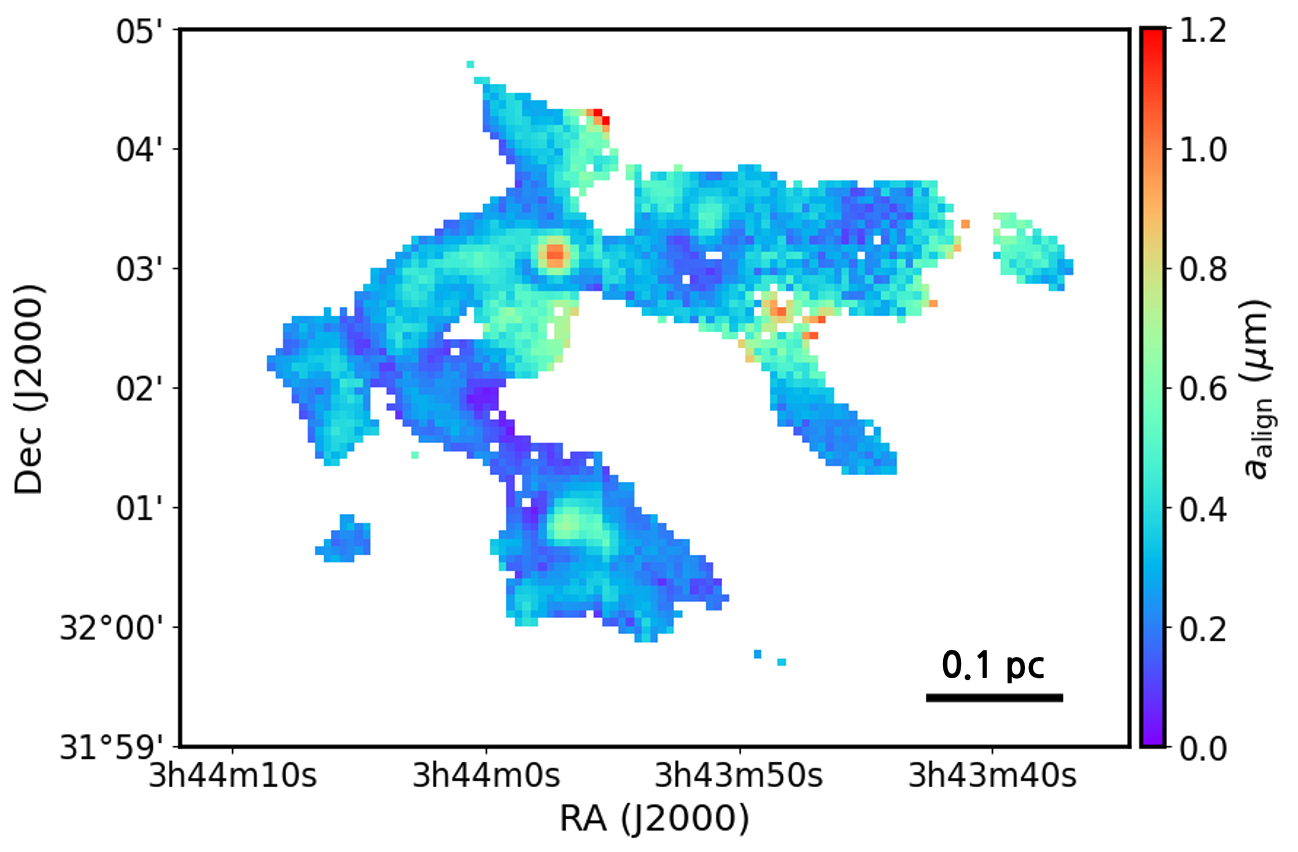}
\caption{The estimated grain alignment size map of the IC 348 star-forming region.
\label{fig:a_align}}
\end{figure}

The slopes of the two core regions ($\alpha = 0.30$ for HH 211 MMS and $\alpha = 0.34$ for IC 348 MMS) are even smaller than that of the filament. Since the radiation from the deeply embedded protostars (HH 211 MMS and IC 348 MMS) appears weak, as indicated by the dust temperature distributions, it is uncertain whether radiation alone is responsible for the low $\alpha$ values in the two core regions. Grain growth in the core regions is another possibility for accounting for the lower slopes. For example, \cite{Pattle_2021} interpreted that the inferred $\alpha$ value of L1689 ($\alpha \sim 0.55$) could be a sign of grain growth since a critical grain size in the region, the minimum size at which grains can be aligned by radiation, is larger than the maximum grain size in the diffuse ISM (0.25-0.3 $\mu$m) \citep{Mathis_1977}. \cite{Hoang_2021} formulated a critical grain size or alignment size ($a_{\mathrm{align}}$) as follows:

\begin{eqnarray}
a_{\mathrm{align}} \simeq &0.055& \hat{\rho}^{-1/7} \left(\frac{\gamma_{-1}}{n_{3}T_{1}}\right)^{-2/7}\left(\frac{\bar{\lambda}}{1.2 \hspace{0.1cm} \mu \mathrm{m}}\right)^{4/7} \nonumber \\
&\times& \left(\frac{T_{d}}{16.4 \hspace{0.1cm} K}\right)^{-12/7} \mu \mathrm{m},
\label{eq:align}
\end{eqnarray}
where $\hat{\rho}=\rho$/(3 g cm$^{-3}$) is the dust mass density, $\gamma_{-1}=\gamma$/0.1 is the anisotropy of the radiation field, $n_{3}=n_{\mathrm{H}}$/($10^{3}$ cm$^{-3}$) is the gas density, $T_{1}=T_{\mathrm{gas}}$/10 K is the gas temperature, $\bar{\lambda}$ is the mean wavelength of the stellar radiation field, and $T_{d}$ is the dust temperature. We use $\gamma_{-1} = 3$ based on the finding of \cite{Bethell_2007}, who reported $\gamma \sim 0.3$ in molecular clouds. The mean incident wavelength in diffuse ISM is about 1.2 $\mu$m \citep{Draine_2007}. According to \cite{Hoang_2021}, the mean wavelength increases in molecular clouds due to attenuation. Therefore, we use $\bar{\lambda}=2$ $\mu$m \citep{Hoang_2021}. Assuming $\hat{\rho} = 1$, $\gamma_{-1}=3$, $\bar{\lambda}=2$ $\mu$m, and $T_{\mathrm{gas}}=T_{\mathrm{dust}}$, we estimated the alignment size in IC 348 using the measured number density and dust temperature in Section \ref{sec:magnetic field strength}. The distribution of alignment size in IC 348 is shown in Figure \ref{fig:a_align}. The alignment size is smaller in the filament and larger in the core regions. By averaging the alignment size in the defined areas, alignment sizes for HH 211 MMS, IC 348 MMS, and the filament are estimated to be 0.39 $\mu$m, 0.57 $\mu$m, and 0.31 $\mu$m, respectively. Since the dust grains are well aligned in the cores, as indicated by the lower power index, a significant portion of the dust grains should be larger than the measured alignment size. Therefore, the maximum grain sizes in the core regions are expected to be micron sizes or larger.

\subsection{Transition from Magnetically Subcritical to Supercritical Condition} \label{subsec:transition}

The measured mass-to-flux ratios are 0.45-2.20 for HH 211 MMS, 0.63-2.76 for IC 348 MMS, and 0.33-1.50 for the filament. Note that these values are upper limit, as only the plane-of-sky magnetic field components are considered. Our results suggest that the filament is likely to be subcritical, and the transition from subcritical to supercritical conditions may occur at the core scale. This is consistent with the recent picture that low-mass stars form in nearly magnetically transcritical environments \citep{Pattle_2023}. For instance, \cite{Karoly_2023} argued that supercritical cores form inside subcritical envelopes in L43 molecular clouds. \cite{Hwang_2021} also found that the magnetically supercritical two clumps in the OMC-1 region are embedded in magnetically subcritical filaments. However, \cite{Ching_2022_2} argued that magnetically supercritical condition is already satisfied during the formation of molecular gas clouds in the L1544 core from Zeeman measurements of HI narrow self-absorption (HINSA) lines. In this case, mass accumulation along field lines or turbulent magnetic reconnection might play more significant roles than ambipolar diffusion in reducing magnetic flux to account for early transition \citep{Ching_2022_2}. The discrepancies between observations could be attributed to the geometric effects, different star-forming cloud conditions, and the uncertainties in measuring the physical quantities, such as magnetic field strengths. To evade these observation uncertainties, \cite{Yin_2021} and \cite{Priestley_2022} post-processed 3D non-ideal MHD simulations and utilized a chemical network to constrain a magnetic criticality in star-forming cores. They suggested that star-forming cores form as initially magnetically subcritical conditions, consistent with the results of our study that star-forming cores form out of a subcritical parent filament. Nevertheless, further studies on the three-dimensional structure of the magnetic field are needed to estimate accurate values of field strengths and, therefore, to constrain when and how magnetically subcritical clouds transition to supercritical star-forming cores to eventually form a star.

\subsection{Stability of the Two Cores} \label{subsec:stability}

The turbulence to magnetic energy ratios are 0.25-2.00 for HH 211 MMS and 0.53-4.25 for IC 348 MMS. This result indicates that the relative strength of the turbulence to the magnetic field tends to be stronger for IC 348 MMS than HH 211 MMS. Interferometric observations toward the two cores revealed a single source driving an outflow in HH 211 MMS, whereas three sources comprise a multiple system in IC 348 MMS \citep{Rodr_guez_2014, Palau_2014}. Since turbulence promotes fragmentation when collapsing to a smaller scale, and magnetic fields have the opposite effect \citep{Hennebelle_2019}, the energy balance at the core scale could possibly explain the different configurations inside the two cores. For HH 211 MMS, strong magnetic fields lead to a single source, while for IC 348 MMS, turbulence drives fragmentation, resulting in a protostellar system with multiple sources \citep{Tang_2019,Chung_2023}.

\section{Conclusions} \label{sec:conclusion}

In this paper, we studied the magnetic field properties of the IC 348 star-forming region as part of the BISTRO survey. We used the 850 $\mu$m polarization data obtained with the POL-2 polarimeter on the JCMT to infer the magnetic field of the region. The magnetic field orientations in IC 348 tend to be perpendicular rather than parallel to the filamentary structure. This result supports the suggestion of \cite{Andr__2014} that local magnetic fields regulate filaments in molecular clouds and is consistent with previous studies \citep[e.g.,][]{Ward_Thompson_2017,Kwon_2022}.

We measured the power-law index of the relation between the polarization fraction and total intensity using two independent methods: the power-law and the mean of the Rice distribution methods. We found that the inferred power indices are larger for the power-law method compared to the Ricean-mean method, as expected by \cite{Pattle_2019}. The Ricean-mean method provides power-law indices of 0.59, 0.30, and 0.34 for the filament, HH 211 MMS, and IC 348 MMS, respectively. This result indicates that dust grains are well aligned even in high-density regions, justifying the use of polarization observations to infer the region's magnetic field. Relatively low power-law index values in the protostellar cores could be attributed to grain growth to micron size or larger in the cores.

The magnetic field strengths of HH 211 MMS, IC 348 MMS, and the filament were measured using the DCF method and the ST modification suggested by \cite{Skalidis_2021}. The measured mass-to-flux ratios are 0.45-2.20 and 0.63-2.76 for HH 211 MMS and IC 348 MMS, respectively, while the ratios for the filament is 0.33-1.50. These results are upper limits, as only the plane-of-sky magnetic fields are considered. Thus, our results suggest that the filament is likely to be magnetically subcritical, where the magnetic field may support the gravitational collapse on the filament scale ($\sim$0.5 pc). The transition from subcritical to supercritical conditions may occur at the core scale ($\sim$0.05 pc).

Lastly, we studied the energy balances of the two core regions. We measured the gravitational, magnetic, thermal, and turbulent energies of the two cores. The turbulence with respect to the magnetic field is likely to be stronger in IC 348 MMS than in HH 211 MMS, which could possibly explain the different configurations inside the two cores: a single protostellar system in HH 211 MMS and multiple protostellar systems in IC 348 MMS.

\vspace{5mm}

We are grateful to the anonymous referee for helpful comments. This work was supported by the National Research Foundation of Korea (NRF) grant funded by the Korea government (MSIT) (RS-2024-00342488). K.P. is a Royal Society University Research Fellow, supported by grant number URF\textbackslash R1\textbackslash211322. F.P. acknowledges support from the Spanish MICINN under grant numbers PID2022-141915NB-C21.

The James Clerk Maxwell Telescope (JCMT) is operated by the East Asian Observatory on behalf of The National Astronomical Observatory of Japan; Academia Sinica Institute of Astronomy and Astrophysics; the Korea Astronomy and Space Science Institute; the National Astronomical Research Institute of Thailand; Center for Astronomical Mega-Science (as well as the National Key R\&D Program of China with No. 2017YFA0402700). Additional funding support is provided by the Science and Technology Facilities Council of the United Kingdom and participating universities and organizations in the United Kingdom, Canada and Ireland. Additional funds for the construction of SCUBA-2 were provided by the Canada Foundation for Innovation. The JCMT has historically been operated by the Joint Astronomy Centre on behalf of the Science and Technology Facilities Council of the United Kingdom, the National Research Council of Canada and the Netherlands Organisation for Scientific Research. The POL-2 data used in this paper were taken under project code M17BL011, and the HARP data were taken under project code M06BGT02. 

The authors wish to recognize and acknowledge the very significant cultural role and reverence that the summit of Maunakea has always had within the indigenous Hawaiian community. We are most fortunate to have the opportunity to conduct observations from this mountain.

\vspace{5mm}

$Facility$: James Clerk Maxwell Telescope (JCMT).

$Software$: Starlink \citep{Currie_2014}, SMURF \citep{Chapin_2013}.

\bibliographystyle{aasjournal}
\bibliography{mybib}

\end{document}

%% file: authorlist.tex
\author[0009-0004-9279-780X]{Youngwoo Choi}
\affiliation{Department of Physics and Astronomy, Seoul National University, Seoul 08826, Republic of Korea}

\author[0000-0003-4022-4132]{Woojin Kwon}
\affiliation{Department of Earth Science Education, Seoul National University, 1 Gwanak-ro, Gwanak-gu, Seoul 08826, Republic of Korea}
\affiliation{SNU Astronomy Research Center, Seoul National University, 1 Gwanak-ro, Gwanak-gu, Seoul 08826, Republic of Korea}
\affiliation{The Center for Educational Research, Seoul National University, 1 Gwanak-ro, Gwanak-gu, Seoul 08826, Republic of Korea}

\author[0000-0002-8557-3582]{Kate Pattle}
\affiliation{Department of Physics and Astronomy, University College London, WC1E 6BT London, UK}

\author{Doris Arzoumanian}
\affiliation{Division of Science, National Astronomical Observatory of Japan, 2-21-1 Osawa, Mitaka, Tokyo 181-8588, Japan}

\author[0000-0001-7491-0048]{Tyler L. Bourke}
\affiliation{SKA Observatory, Jodrell Bank, Lower Withington, Macclesfield SK11 9FT, UK}
\affiliation{Jodrell Bank Centre for Astrophysics, School of Physics and Astronomy, University of Manchester, Oxford Road, Manchester, UK}

\author[0000-0003-2017-0982]{Thiem Hoang}
\affiliation{Korea Astronomy and Space Science Institute, 776 Daedeokdae-ro, Yuseong-gu, Daejeon 34055, Republic of Korea}
\affiliation{University of Science and Technology, Korea, 217 Gajeong-ro, Yuseong-gu, Daejeon 34113, Republic of Korea}

\author[0000-0001-7866-2686]{Jihye Hwang}
\affiliation{Korea Astronomy and Space Science Institute, 776 Daedeokdae-ro, Yuseong-gu, Daejeon 34055, Republic of Korea}

\author[0000-0003-2777-5861]{Patrick M. Koch}
\affiliation{Academia Sinica Institute of Astronomy and Astrophysics, No.1, Sec. 4., Roosevelt Road, Taipei 10617, Taiwan}

\author{Sarah Sadavoy}
\affiliation{Department for Physics, Engineering Physics and Astrophysics, Queen{'}s University, Kingston, ON, K7L 3N6, Canada}

\author[0000-0002-0794-3859]{Pierre Bastien}
\affiliation{Centre de recherche en astrophysique du Qu\'{e}bec \& d\'{e}partement de physique, Universit\'{e} de Montr\'{e}al, C.P. 6128 Succ. Centre-ville, Montr\'{e}al, QC, H3C 3J7, Canada}

% ...
